\documentclass[review]{cas-sc}

\usepackage{lineno}                % Line numbering
\usepackage{graphicx}              % Figures and images
\usepackage{subcaption}            % Subfigures (after graphicx)
\usepackage{adjustbox}             % Resize tables/figures

\usepackage{amsmath, amssymb}      % Math environments and symbols
\usepackage[version=4]{mhchem}     % Chemical formulas and reactions

\usepackage{booktabs}              % Professional-looking tables
\usepackage{multirow}              % Merged cells
\usepackage{arydshln}              % Dashed lines in tables

\usepackage{caption}               % Control caption formatting

\usepackage[authoryear]{natbib}    % Author-year citation style

\usepackage[commandnameprefix=ifneeded]{changes}
\usepackage[normalem]{ulem}        % Underline / strikeout (must come AFTER changes)
\usepackage{soul}                  % Highlighting, underlines

\setulcolor{blue}                  % Underline color (for soul)
\setul{0.3ex}{0.8pt}               % Underline depth and thickness

\DeclareCaptionFont{mysingle}{\setstretch{1}}
\def\tsc#1{\csdef{#1}{\textsc{\lowercase{#1}}\xspace}}
\tsc{WGM}\tsc{QE}\tsc{EP}\tsc{PMS}\tsc{BEC}\tsc{DE}

\begin{document}
%\linenumbers                          % line numbers
\let\WriteBookmarks\relax
\def\floatpagepagefraction{1}
\def\textpagefraction{.001}

% ===============================
% TITLE AND AUTHORS
% ===============================
\title[mode=title]{Identification and Characterization of the Topside Bulge of the Venusian Ionosphere}

\author[1,2]{Satyandra M. Sharma}
\ead{smsharma@prl.res.in}

\author[1]{Varun Sheel}
\cormark[1]
\ead{varun@prl.res.in}

\author[3]{Martin Pätzold}
\ead{martin.paetzold@uni-koeln.de}

% ===============================
% AFFILIATIONS
% ===============================
\affiliation[1]{organization={Planetary Sciences Division, Physical Research Laboratory, Navarangpura},
               addressline={Ahmedabad 380009}, 
               country={India}}

\affiliation[2]{organization={Indian Institute of Technology Gandhinagar  Palaj},
               addressline={Gandhinagar 382355}, 
               country={India}}

\affiliation[3]{organization={Rheinisches Institut für Umweltforschung, Abteilung Planetenforschung, Universität zu Köln},
               addressline={Cologne}, 
               country={Germany}}

% Corresponding authors
\cortext[cor1]{Corresponding author: Varun Sheel}

\begin{abstract}
Venus, in the absence of an intrinsic magnetic field, undergoes a direct interaction between its ionosphere and the solar wind. Previous missions, including \textit{Mariner}, \textit{Venera}, and the \textit{Pioneer Venus Orbiter} (\textit{PVO}), reported a recurring localized increase in electron density, often termed a "bulge," at altitudes between $160$ and $200~km$. This study investigates this topside bulge using over $200$ dayside electron density profiles derived from the \textit{Venus Radio Science experiment (VeRa)} onboard the \textit{Venus Express (VEX)}. We employ an automated, gradient-based classification algorithm to provide a quantitative and reproducible method for identifying and categorizing the bulge morphology into three types. Type 1 profiles exhibit a distinct secondary peak above the main V2 layer. Type 2 profiles display a shoulder-like feature near the bulge altitude. Type 3 bulges are not visually apparent but can be identified through residuals obtained after subtracting a Chapman layer fit to the V2 peak. The bulge is detected in over $80\%$ of the analyzed profiles, with a higher occurrence during periods of low solar activity and at lower solar zenith angles ($SZAs$). Type 1 morphologies are only observed at low latitudes (within $\pm 40^\circ N$). The peak altitude of the bulge negatively correlates with $SZA$, suggesting that thermospheric cooling toward the terminator significantly influences the bulge altitude. The occurrence patterns and morphological characteristics indicate that the bulge is likely influenced by external drivers, such as solar wind interaction, rather than being solely a result of local photochemical processes.
\end{abstract}

\begin{graphicalabstract}
\end{graphicalabstract}

\begin{highlights}
\item This work presents the first in-depth analysis of the Venusian topside ionospheric bulge and its systematic characterization. 
\item The bulge is observed in more than $80\%$ of the VeRa electron density profiles.
\item Its occurrence is more frequent during periods of low solar activity and at smaller solar zenith angles.
\item The dependence of bulge peak density, altitude, and occurrence rate indicates that its origin cannot be explained primarily by photochemical processes.
\end{highlights}

\begin{keywords}
Venus\\
Ionosphere\\
Bulge\\
Radio Science\\
Venus Express\\
\end{keywords}

\maketitle
%%%%%%%%%%%%%%%%%%%%%%%%%%%%%%%%%%%%%%%%%%%%%%%
%
% introduction
%
%%%%%%%%%%%%%%%%%%%%%%%%%%%%%%%%%%%%%%%%%%%%%%%
\section{Introduction}
\label{Introduction}

The Venusian ionosphere was first detected through the Radio Occultation (RO) experiment on \textit{Mariner 5} in 1967, which provided the first electron density profile showing a peak of $5 \times 10^5$ to $6 \times 10^5~cm^{-3}$ near $142~km$ altitude at a solar zenith angle $33.3^\circ$, along with a minor layer about $15~km$ below the main peak \citep{Kliore1967}. Subsequent missions, including \textit{Mariner 10}, \textit{Venera 9/10}, \textit{Venera 15/16}, \textit{Pioneer Venus Orbiter} (\textit{PVO}), and \textit{Magellan}, further expanded our knowledge of the Venusian ionosphere through dayside radio sounding profiles. 

Pre-1995 datasets included single profiles from \textit{Mariner 5} and \textit{Mariner 10}, multiple profiles from the \textit{Venera} missions, a large volume from \textit{PVO} (148 profiles recorded between 1979 and 1989) \citep{kliore1979initial, kliore1991solar} and 14 profiles from \textit{Magellan} between 1992 and 1994 \citep{steffes1994radio}. The most extensive dataset prior to 2000 was created by PVO’s Orbiter Radio Occultation (ORO) experiment \citep{kliore1979initial}. A limited number of original profiles from the first occultation season is available from the PDS \citep{Withers2022PVO}, but those are altitude biased by many kilometers as new reprocessing suggests \citep{Patzold2022EPSC, Patzold2022AGU, Patzold2023AGU, Oschlisniok2024EGU, Patzold2024EPSC}.

In the post-2000 era, the \textit{Venus Express} (VEX) mission, which orbited Venus from 2006 to 2014, significantly enhanced the available dataset by collecting over 900 electron density profiles using the radio occultation technique \protect\citep{hausler2006radio, Patzold2007}. Recently, the Akatsuki mission also contributed to the investigation of the Venus ionosphere from 2016 to 2024, using radio occultation experiment \citep{imamura2017initial, tripathi2023venusian}. Scientific analysis is still under progress.

%%%%%%%%%%%%%%%%%%%%%%%%%%%%%%%%%%%%%%%%%%%%%%%
% Structure of Venus' Ionosphere
%%%%%%%%%%%%%%%%%%%%%%%%%%%%%%%%%%%%%%%%%%%%%%%
\subsection{Structure of the Venusian ionosphere}
The ionosphere is considered a region of the atmosphere where free thermal ($1eV$) electrons and ions are present \citep{schunk2000ionospheres}. The radio occultation method is only sensitive to the electron population in the ionosphere. Therefore, the electron density distribution is a representation of the total plasma distribution. The dayside Venusian ionosphere is characterized by a dominant ion layer V2 at $135 - 145~km$ with a peak electron density controlled by the solar zenith angle ($SZA$) \citep{Patzold2007, gerard2017aeronomy}. A secondary peak, named as the V1 layer of the ionosphere, is identified in the altitude range of $125 - 130~km$ below the V2 layer \citep{breus1985properties, girazian2015characterization}. The V2 and V1 layers are formed mainly by the photoionization of neutral \ce{CO2} by solar EUV and solar soft X-rays, respectively, creating \ce{CO2+} ion and photoelectrons. Photoelectrons produced by the soft X-rays contribute further to the ion production by the impact ionization, which dominates at the V1 layer \citep{peter2014dayside}. The primarily produced \ce{CO2+} ion is readily converted into \ce{O2+} through ion-neutral charge exchange reactions with atomic \ce{O}, making \ce{O2+} the dominant ion in these layers \citep{kumar1974venus}. The \ce{O2+} ion is mainly lost via dissociative recombination with electrons. The formation of the V2 and V1 layers can be explained within the framework of photochemical equilibrium (PCE), an assumption that remains valid up to altitudes of approximately $170~km$ \citep{mendillo2020ionosphere, martinez2024three}.

\begin{figure}
    \centering
    \includegraphics[width=1\linewidth]{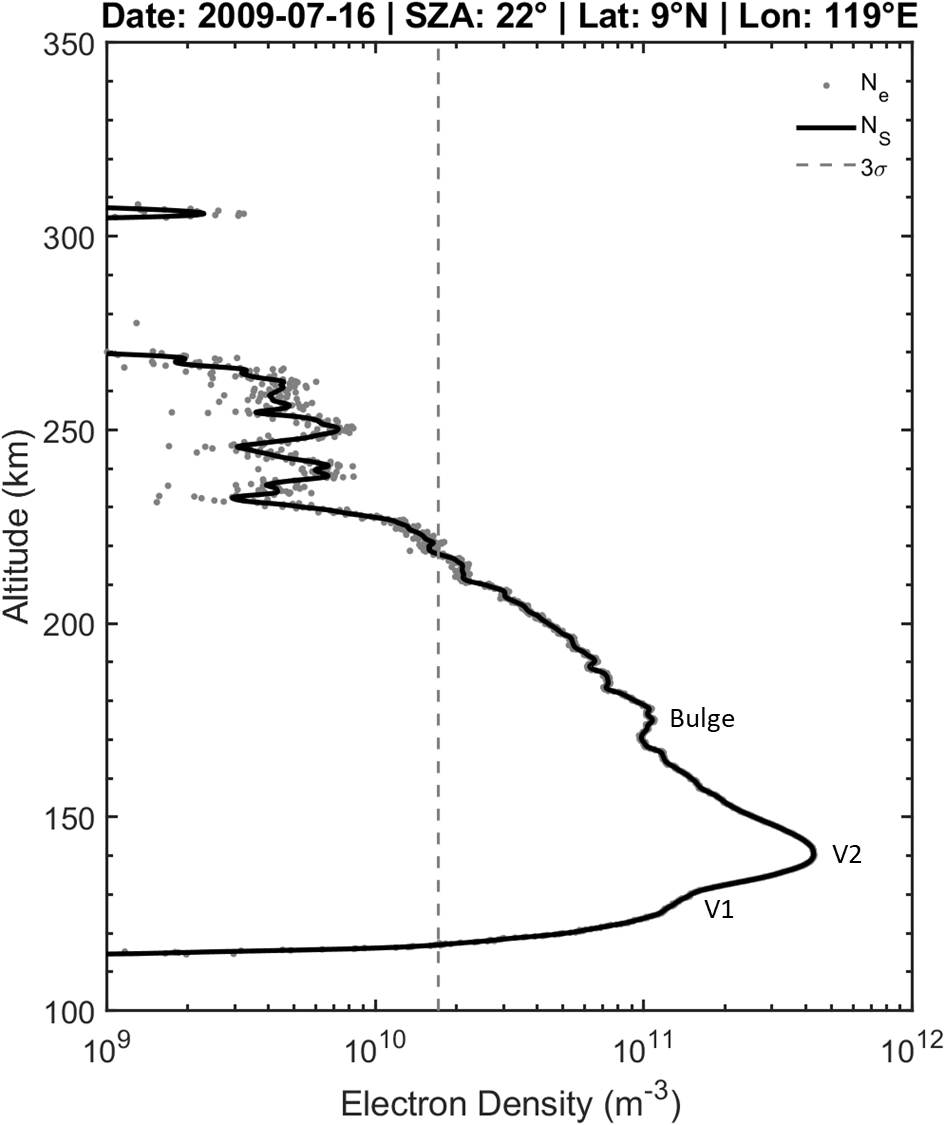}
    \caption{
    A Venus Express VeRa radio occultation daytime electron density profile observed on 16 July 2009. The gray spheres represent the retrieved electron density, $N_{e}$. The black line shows the smoothed profile, $N_{s}$, obtained after 10 successive iterations using Equation~\ref{eq: smoothing}. The vertical light gray line represents the $3\sigma$ uncertainty in the retrieval. Two ionospheric layers are marked: V2, the main peak formed by solar EUV ionization, and V1, formed by soft X-rays and secondary photoelectron impact ionization. The topside bulge located above the V2 layer is investigated in this work. The solar zenith angle during this observation was $22^\circ$.
    }
    \label{Figure_1}
\end{figure}

The photoionization of the atomic oxygen \ce{O} dominates above the $170~km$, forming the \ce{O+} ion, which is the major ion there \citep{fox2001solar, peter2014dayside}. The longer chemical lifetime of the \ce{O+} ion as compared to the transport timescales enables the ion-diffusion process to control the topside ionospheric structure, forming a diffusion-dominated region there \citep{Patzold2007}. A recurrent feature within this region is the appearance of a bulge (Figure~\ref{Figure_1}), which is not reproduced by existing theoretical models of the Venusian ionosphere  \citep{chen1978comprehensive, Patzold2007, peter2014dayside}. %Emergence of this bulge will therefore be indicative of additional processes superimposed on the diffusion-controlled region, which we will investigate further.

At the upper boundary of the ionosphere (at $250~km$ to $400~km$), a sharp electron density gradient is observed over a short altitude range ($30 -50~km$) \citep{gerard2017aeronomy}. This boundary is referred to as the ionopause. The ionopause forms at the altitude where the solar wind dynamic pressure is balanced by the thermal pressure of the ionosphere \cite{elphic1980observations}, thereby separating the solar wind plasma from the ionospheric plasma. However, during periods of high solar wind dynamic pressure, the Venusian ionosphere is often unable to fully withstand the external forcing. Under such conditions, the interplanetary magnetic field can penetrate into the ionosphere, contributing to the total internal pressure and thereby restoring the pressure balance \citep{luhmann1991magnetic}.

%%%%%%%%%%%%%%%%%%%%%%%%%%%%%%%%%%%%%%%%%%%%%%%
% Early studies of bulge
%%%%%%%%%%%%%%%%%%%%%%%%%%%%%%%%%%%%%%%%%%%%%%%
\subsection{Early observations of the topside bulge}
The Mariner-10 spacecraft, during its Venus flyby on February 5, 1974, revealed a bulge-like feature in the electron density profile obtained through radio occultation measurements \citep{Howard1974, fjeldbo1975mariner}. The observed electron density profile exhibited a "gliding staircase" appearance with two distinct ledges, one near $180~km$ and the other near $250~km$ altitude. The lower ledge, termed the "F2 bulge" by \citet{bauer1974venus}, was interpreted as the altitude where photochemical processes involving \ce{O+} ions (production by photoionization and loss through a charge transfer process with \ce{CO2}) are balanced by diffusion. Slightly above the \ce{O+} peak, the solar wind-induced downward transport becomes the controlling factor for the \ce{O+} distribution.

The bulge has been referred to by different terms in the literature, including ``ledge'' \citep{chen1978comprehensive}, ``protrusion'' \citep{gavrik2008inhomogeneous}, ``photodynamical ionopause'' \citep{mahajan1989venus}, and ``bulge'' \citep{Patzold2007, peter2014dayside}.

Observations from the Venera-9 and Venera-10 missions in 1975 revealed the solar zenith angle ($SZA$) dependence of the bulge. This feature exhibited a secondary maximum above the main layer at lower $SZAs$ and became less distinct, transitioning into a "kink"-like structure at higher $SZAs$ \citep{ivanov1979daytime}.

\citet{chen1978comprehensive} developed a two-dimensional model but could not reproduce the 'bulge' or 'ledge' observed in Mariner 10 data. In contrast, \citet{nagy1975model} found that their model could reproduce the bulge if the electron temperature is made to increase from the neutral temperature of $ 350~K$ at $ 178~km$ to about $ 900~K$ at $ 188~km$.  However,  \citet{butler1978photoelectrons} found no evidence of a sharp gradient in electron temperature. 

A model study by \citet{fox2001solar}  conducted under high solar activity shows the presence of a 'bulge' composed of O$^+$ ions formed at around $\sim200~km$ altitude. They attributed the formation of this bulge to the high atomic oxygen \ce{O} density derived from the global empirical model of the Venus Thermosphere (VTS3) by \citet{hedin1983global}. Similarly, \citet{Butler1975Ionosphere} reported a comparable bulge in their model results, concluding that the $190~km$ feature observed in the Mariner 10 data was almost certainly produced by \ce{O+} ions. 

\citet{woo1991magnetization} studied the 148 dayside electron density profiles obtained from orbital radio occultation (ORO) experiments onboard the \textit{PVO}. They found that the occurrence frequency of ledges was high at small $SZAs$ and remained similar during both solar minimum and solar maximum. However, during solar minimum, ledges are observed at higher $SZAs$ as well. This dataset has limited profiles at low $SZA$, particularly during solar minimum, where most observations were confined to $SZAs$ greater than 60°.
%The VeRa dataset provides an excellent opportunity to study the topside bulge of the Venusian ionosphere under conditions of low solar activity, spanning the solar minimum between Solar Cycles 23 and 24.

%%%%%%%%%%%%%%%%%%%%%%%%%%%%%%%%%%%%%%%%%%%%%%%
%
% Objectives
%
%%%%%%%%%%%%%%%%%%%%%%%%%%%%%%%%%%%%%%%%%%%%%%%
\subsection{Objectives}

This study makes use of electron density profiles retrieved from the \textit{Venus Express Radio Science Experiment} (VeRa) \footnote{\textcolor{blue}{The VEX-VeRa data radio occultation set is available on request from the experiment team. Contact co-author P{\"a}tzold under martin.paetzold@uni-koeln.de.}}. Solar wind measurements from \textit{the Analyzer of Space Plasmas and Energetic Atoms - 4} (ASPERA-4) onboard \textit{Venus Express} (VEX) \citep{barabash2007analyser} (Automated Multi-Dataset Analysis (AMDA) available at \url{https://amda.irap.omp.eu/}) has been used to study the influence of solar wind on the topside ionospheric bulge. We also utilize Earth-based solar activity data ($F_{10.7}$) appropriately scaled to the heliocentric position of Venus , to characterize solar activity during the VeRa observations. The primary objectives are as follows:

\begin{itemize}
    \item[\textbf{A.}] \textbf{Identification of the bulge:} 
    There is no universally accepted method for defining the bulge in the Venusian ionosphere. Nevertheless, once identified, it can provide valuable information about the nature of its occurrence, timing of formation, and the conditions that favor its evolution.

    \item[\textbf{B.}] \textbf{Characterization of the bulge by altitude ($h_{m,\mathrm{bulge}}$) and density ($N_{m,\mathrm{bulge}}$):} 
    This objective focuses on quantifying the bulge by determining its peak altitude ($h_{m,\mathrm{bulge}}$) and peak electron density ($N_{m,\mathrm{bulge}}$). The analysis also explores how the properties of the bulge vary with $SZA$ and solar activity, providing insight into the controlling factors behind its manifestation.

    \item[\textbf{C.}] \textbf{Morphology of the bulge:} 
    The morphology of the layer is indicative of the relative roles of ongoing physical processes depending on the background conditions. 
\end{itemize}

Section \ref{Data and Methodology} describes the VeRa electron density profiles, the algorithm developed to detect the bulge, and the solar flux data used to characterize the Solar activity level during the observations. Section \ref{Results} presents an in-depth analysis of the results, followed by a discussion in Section \ref{Discussion}. The conclusions are provided in Section \ref{Conclusions}. 

%%%%%%%%%%%%%%%%%%%%%%%%%%%%%%%%%%%%%%%%%%%%%%%%%
%
% Data and Methodology
%
%%%%%%%%%%%%%%%%%%%%%%%%%%%%%%%%%%%%%%%%%%%%%%%%%
\section{Data and Methodology}
\label{Data and Methodology}
%%%%%%%%%%%%%%%%%%%%%%%%%%%%%%%%%%%%%%%%%%%%%%%%%%
% Electron density
%%%%%%%%%%%%%%%%%%%%%%%%%%%%%%%%%%%%%%%%%%%%%%%%%%
Electron density profiles retrieved from the radio occultation (RO) experiment onboard \textit{Venus Express} (VEX) form the basis of this study. The spacecraft was launched on 9 November 2005 and was inserted into orbit around Venus on 11 April 2006. The RO technique is a well-established method for retrieving vertical profiles of planetary atmospheres and ionospheres, based on the principles of geometrical optics. It was first developed in the 1960s \citep{FjeldboRO65}, and has since been applied extensively to Venus, Mars, and other planetary bodies \citep{Howard1974, hausler2006radio, patzold2016comment, patzold2016mars}. During an RO experiment, a spacecraft transmits coherent radio signals toward Earth as it moves behind the planet’s limb as seen from the Earth. These signals pass through the planet’s atmosphere and ionosphere, where they are refracted due to spatial gradients in the refractive index. In the neutral atmosphere, refraction is governed by pressure, temperature, and neutral number density, while in the ionosphere, it is driven by the electron density. The resulting change in the raypath leads to a phase shift in the radio carrier, which is observed as a frequency shift in the ground station antenna receiver. By subtracting a predicted geometric Doppler shift---based on spacecraft ephemeris and precise orbit determination assuming that the planet has no atmosphere---from the observed signal, one obtains the frequency residuals caused solely by atmospheric and ionospheric bending. This frequency shift is directly proportional to the bending angle $\alpha(r)$ of the ray, which is then related to the refractive index $n(r)$ via an Abel transform inversion \citep{FjeldboRO65, patzold2016comment}. 

From the refractive index, one can obtain the neutral($N(r)$) and electron ($N_e(r)$) number densities at the distance $r$ from the center of the planet \citep{FjeldboRO65,patzold2004mars, Patzold2007, patzold2016comment}.

\begin{figure}
    \centering
    \includegraphics[width=1\linewidth]{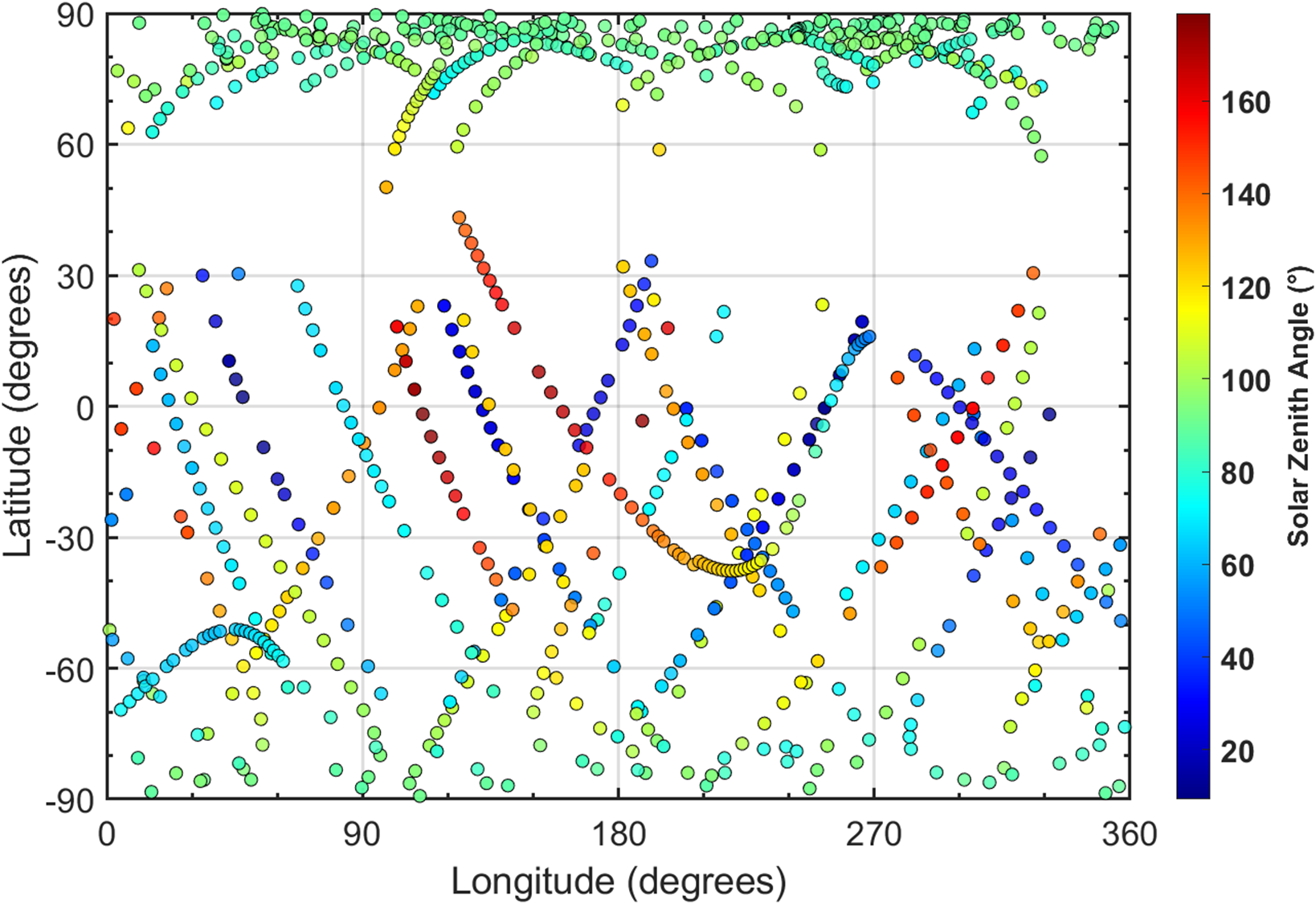}
    \caption{Spatial distribution of all VeRa profiles on a Venus latitude--longitude map. 
    The gap between $40^\circ$ and $60^\circ$ latitude arises from the orbital geometry of Venus Express (VEX), which follows a highly elliptical orbit. Because the pericenter of VEX is close to the north pole, most profiles at high northern latitudes do not extend above $350~km$ and are therefore excluded from our analysis.}
    \label{Figure_2}
\end{figure}

The \textit{Venus Express Radio Science Experiment (VeRa)} \citep{hausler2006radio, Patzold2007} was designed to perform radio occultation studies of the Venusian ionosphere and neutral atmosphere using two downlink signals at X and S bands \citep{hausler2006radio}. Vertical profiles of ionospheric electron density can be derived from the analysis of either a single radio signal  (typically the X-band) or from the differential Doppler measurements using both X- and S-band signals \citep{patzold2016comment, patzold2016mars}. Differential Doppler profiles provide the true electron density distribution in the ionosphere, and we prioritize their use when available. They are preferred because the two-frequency technique eliminates contaminants arising from residual errors in the spacecraft and ground station relative velocities. When the second frequency is unavailable, the electron density profile derived from a single frequency may be biased, particularly at higher altitudes (topside), and should therefore be interpreted with great caution. The VeRa dataset constitutes the most extensive collection of electron density profiles of the Venusian ionosphere to date, providing near-global coverage (Figure~\ref{Figure_2}). The observations span a wide range of solar zenith angles ($SZA$), thereby enabling the investigation of both dayside and nightside ionospheric structures. Furthermore, the dataset covers solar cycle 23 and the rising phase of solar cycle 24, enabling investigations of solar cycle–dependent variability in the Venusian ionosphere.

\subsection{Methodology to identify the topside bulge}
\label{Electron density}
\begin{figure}[ht]
    \centering
    \includegraphics[width=0.8\textwidth]{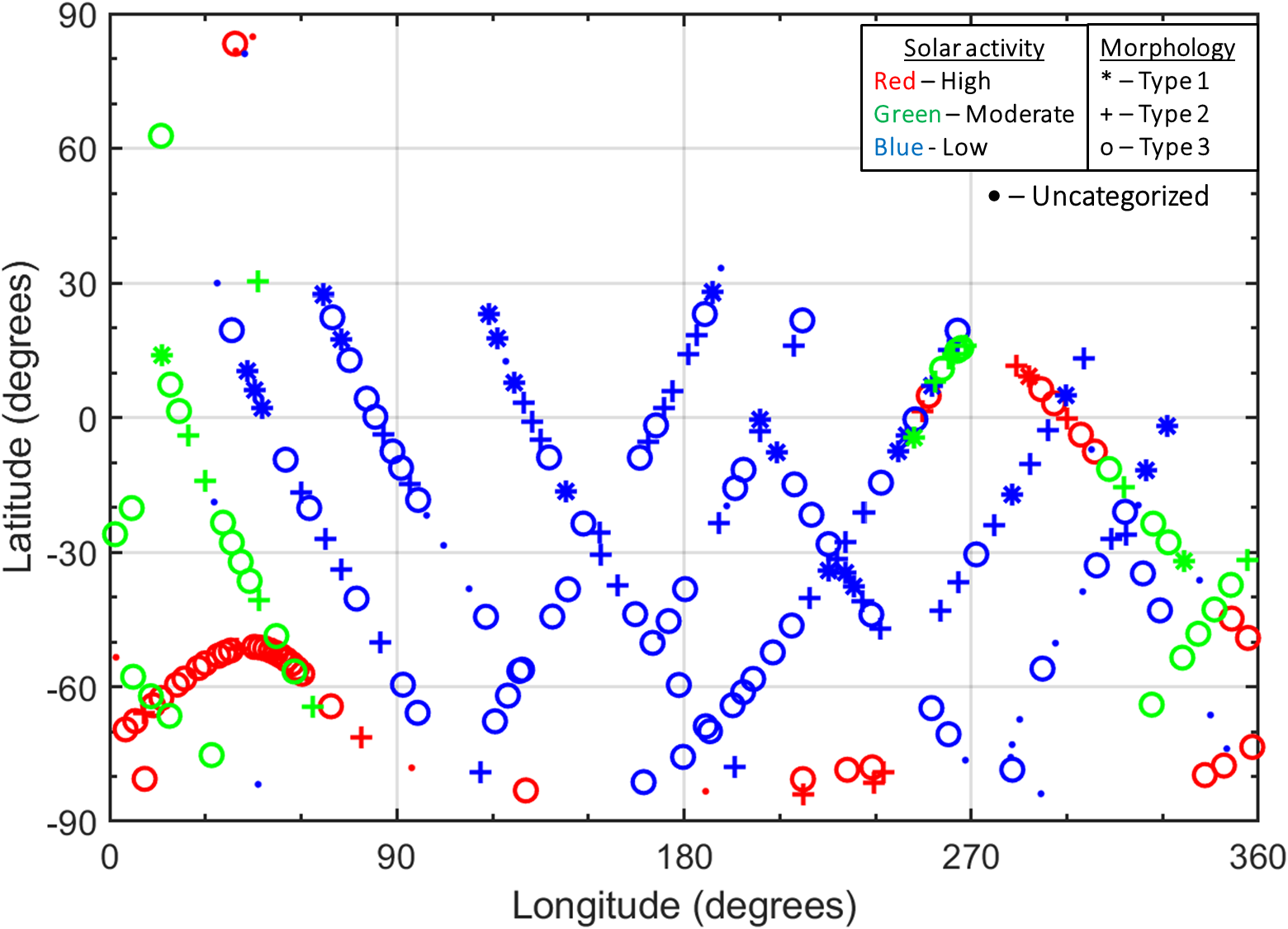}
    \caption{Spatial distribution of selected 234 electron density profiles in the Venusian ionosphere used in this study. Observations are primarily concentrated in the equatorial and southern hemisphere regions. High, moderate, and low solar activity levels are indicated by red, green, and blue, respectively. Symbols represent morphologies: star for Type 1, cross for Type 2, open circle for Type 3, and closed circle for Type 4.}
    \label{Figure_3}
\end{figure}

For the identification of the bulge, only dayside electron density profiles with $SZA$ $\leq 85^{\circ}$ were considered. Profiles near the terminator were excluded, since the Chapman function exhibits significant deviations at larger $SZAs$ \citep{peter2014dayside,peter2018small,mukundan2022m3}. Furthermore, only those profiles that extended to altitudes above $350~\mathrm{km}$ were retained for the analysis. The altitude criterion eliminates profiles whose pre-occultation baseline is insufficient for adequate calibration. Because the Venus Express (VEX) spacecraft had a highly elliptical orbit with its pericenter near the north pole, most profiles in the northern hemisphere did not meet the altitude criterion and were therefore excluded \citep{peter2025variability}. As a result, our analysis is mainly restricted to the low latitudes of the northern hemisphere and to the southern hemisphere (Figure~\ref{Figure_3}). The study focuses on dayside electron density profiles because the Venusian ionosphere there is primarily controlled by solar EUV-driven photochemistry and follows a Chapman-like behavior, enabling reliable modeling of the background profile and extraction of the bulge contribution. Nightside profiles, dominated by plasma transport rather than photoionization, do not exhibit such behavior, and their inclusion would bias the analysis and hinder meaningful comparisons.

If, after application of these criteria, a differential Doppler profile is available for a given occultation, then we use it. Otherwise, we used the X-band profile cautiously. However, some differential Doppler profiles had large data gaps, did not extend down to the V1 layer, or had uncharacteristic electron densities likely caused by signal tracking errors during data recording. These profiles were replaced by their corresponding X-band versions when available. In addition to this, we removed the profiles whose noise level \(\sigma_{\text{noise, av}} > 0.5 \times N_{m, V2}\). The differential Doppler profiles have an average noise in electron density of $1.5 \times 10^{10}~m^{3}$ and the X-band only profiles have a smaller average noise $1.0 \times 10^{9}~m^{3}$ \citep{girazian2015characterization}.

 After application of these criteria, we are left with $234$ electron density profiles with $SZAs$ between $9.3^{\circ}$ and $84.9^{\circ}$ (Figure~\ref{Figure_3}). The profiles were obtained between 15 July 2006 and 13 February 2014, during which the solar activity went from the deep solar minimum of Solar Cycle 23 to the rising solar activity phase of Solar Cycle 24. We smoothed each electron density profile by the following procedure in order to reduce any small-scale noise in the profile \citep{peter2014dayside}:
\begin{equation}
    \ N_{s}(h_i) = \frac{2N_e(h_i) + \sum_{\substack{j \neq 0 \\ j = -2}}^2 \frac{0.5 \left| h_{i-1} - h_{i+1} \right|}{\left| h_{i-j} - h_i \right|} N_e(h_{i+j})}{2 + \sum_{\substack{j \neq 0 \\ j = -2}}^2 \frac{0.5 \left| h_{i-1} - h_{i+1} \right|}{\left| h_{i-j} - h_i \right|}} \
\label{eq: smoothing}
\end{equation} 
 where $N_{s}(h_i)$ is the smoothed electron density at altitude $h_i = r - R_{Venus}$; where $R_{Venus}$ is the radius of Venus $(6051.8~km)$. This iteration is applied 10 times to reduce the small-scale noise. Iterations beyond ten produced negligible changes, indicating convergence; thus, ten iterations were adopted to effectively suppress inversion noise while preserving the physical morphology of the profiles. The typical vertical resolution of the smoothed profiles is $1~km$.

An automated detection routine was developed to identify the bulge feature above the V2 peak. To isolate this structure, the contribution of the V2 layer was first removed by applying a Chapman fit in the vicinity of the V2 peak, thereby extracting the residual topside ionospheric profile. The Chapman fit is the most accurate fit to the ionospheric layer in an idealized isothermal ionosphere controlled by the photochemistry only. \citep{chapman1931absorption,schunk2000ionospheres, fallows2015observational, girazian2015characterization, mayyasi2018sporadic, peter2018small}. It adequately fits observations and can provide reproducible and reliable predictions of ionospheric profile structure. We first locate the V2 layer peak electron density $N_{m, V2}$ and peak altitude $h_{m, V2}$ in an ionospheric profile. A Chapman fit is generated using $N_{m,V2}$ and $h_{m, V2}$ as follows:
 \begin{equation}
     N_{chap, V2}(h, \chi) = N_{m, V2}^{0} \exp\left(\frac{1}{2} \left( 1 - \frac{h - h_{m,V2}^{0}}{H} - Ch^{*}(X_{p},\chi) \exp\left(-\frac{h - h_{m,V2}^{0}}{H}\right) \right) \right)
\label{eq:V2_Chapman}
 \end{equation}
 where $N_{chap, V2}$ is the electron number density in $m^{-3})$ generated by the Chapman fit at the altitude $h$ (in $km$) and $SZA$ $\chi$. $N_{m, V2}^{0}$ (in $m^{-3})$) is the peak electron density of the V2 layer, and $h_{m, V2}^{0}$ is the corresponding peak altitude (in $km$) at $\chi = 0^\circ$. $H$ is the neutral scale height (in $km$) of the atmosphere at  $h_{m, V2}$ (typically $5-6~km$ at the V2 peak which is consistent with the VTS 3 model \citep{hedin1983global}). $ Ch^{*}(X_{p},\chi)$ is the Chapman function approximation for grazing incidence angles derived by \citet{smith1972numerical}, where $X_p = r_{p}/H$ with $r_p$ being the radial distance of the point from the center of the planet. The Chapman fit was performed on three different subsets of the profile around the $N_{m, V2}$ \citep{PETER2023115565}. In the first criterion, the fitting range included the region above and below the$N_{m, V2}$ where the electron density remained greater than 75\% of the $N_{m, V2}$. In the second criterion, the fit range was selected between the altitudes where the gradient of the smoothed electron density profile, $(N_{s})'$, reaches its local extrema near the $N_{m, V2}$. Specifically, this includes the altitude of the maximum $(N_{s})'$ altitude just above the peak ($h_{BL,2}$) and the minimum $(N_{s})'$  altitude just below it ($h_{V1,2}$). In the third criterion, the boundaries were chosen where $(N_{s})'$ decreases to 70\% of its local extremum values. That is, $h_{V1,1}$ marks the altitude below the $h_{V1, 2}$ where the $(N_{s})'$  reduces to 70\% of its maximum, and $h_{BL,1}$ marks the altitude above the $h_{BL,2}$ where the $(N_{s})'$ increases to 70\% of its minimum \citep{peter2021lower, PETER2023115565}. Here, $h_{V1,1}$ and $h_{V1,2}$ define the V1 layer boundaries in a region minimally influenced by the V2 peak, while $h_{BL,1}$ and $h_{BL,2}$ define the boundaries of the potential bulge feature located above the V2 peak. Among these three fits, the Chapman fit yielding the smallest average deviation from the observed electron density ($N_e$) in the selected region was considered the final fit \citep{peter2025variability}. The resulting Chapman profile, $N_{\mathrm{chap}, V2}$, was then subtracted from the smoothed electron density profile, $N_{s}$, to obtain the residual profile, $N_{\mathrm{res}}$.

The presence of a bulge causes a deviation from the typical Chapman-like electron density distribution, and this deviation is more evident in the residual profile, $N_{\mathrm{res}}$. In most profiles, $N_{\mathrm{res}}$ exhibits a distinct maximum within the same altitude range. The maximum density, $N_{m, \mathrm{BL}}$, and corresponding altitude, $h_{m, \mathrm{BL}}$, of $N_{\mathrm{res}}$ are used to characterize the bulge in the Venusian ionosphere. The bulge width, $W_{bulge}$, is defined as the altitude range over which $N_{\mathrm{res}}$ remains greater than 80\% of $N_{m, \mathrm{BL}}$.  A detected bulge is considered valid if its width lies between $5~\mathrm{km}$ and $40~\mathrm{km}$ and its peak density satisfies $N_{m, \mathrm{BL}} < 6\sigma$ \citep{peter2025variability}. This width criterion helps exclude broader variations arising from diffusive extensions and other small-scale perturbations in the ionosphere.

\begin{figure}[ht]
    \centering
    \includegraphics[width=0.8\linewidth]{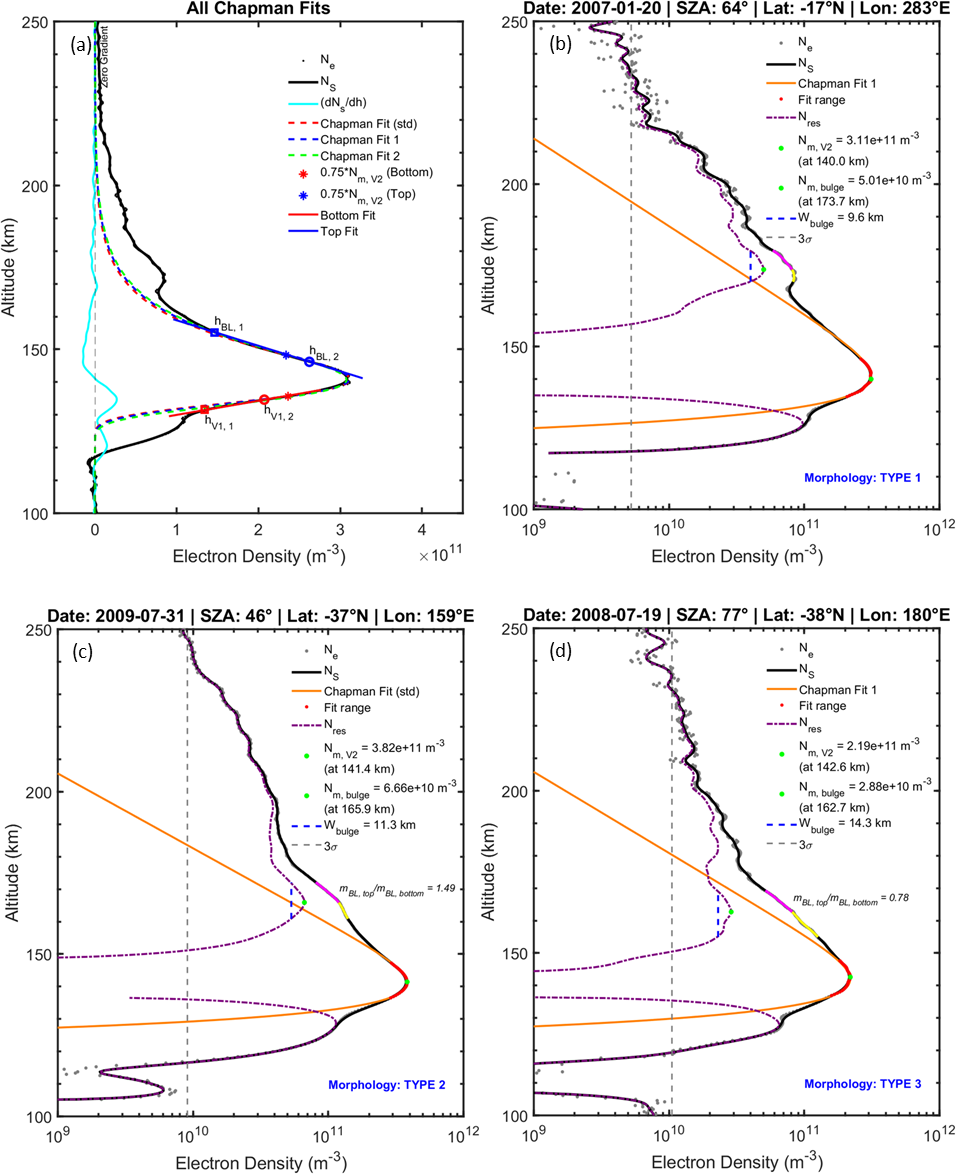}
    \caption{Examples of the three bulge morphologies identified in VeRa electron density profiles using the Chapman fit method (Section ~\ref{Electron density}:
    (a) Illustration of different boundary selections for Chapman fitting; (b) Type 1 — The bulge appears as a distinct secondary peak located above the main V2 layer; (c) Type 2 — The bulge shows as a shoulder-like enhancement near the V2 peak; and (d) Type 3 — No visually distinct bulge is present, but the residual electron density ($N_{res}$) obtained after subtracting a Chapman fit ($N_{chap, V2}$) from the smoothed profile ($N_{s}$) indicates a layer above the V2 peak. Profiles that do not contain a valid bulge based on the defined criteria are not categorized. Here, $W_{\mathrm{bulge}}$ represents the width of the bulge, and $m_{\mathrm{BL,top}}$ and $m_{\mathrm{BL,bottom}}$ denote the average gradients of the $N_{S}$ above and below $N_{m,\mathrm{BL}}$, respectively, within the width of the bulge (see Section~\ref{Electron density}).
    } 
    \label{Figure_4}
\end{figure}

Each valid bulge is further categorized into one of three morphology types based on the best fit to the electron density profile (Figure ~\ref{Figure_4}). In contrast to earlier categorizations of bulge morphology based solely on visual inspection \citep{mayyasi2018sporadic, mukundan2022m3}, we developed an automated routine to identify the bulge morphology using the gradient of the smoothed electron density profile, $(N_{s})' = dN_{s}/dh$. Building on the definition of a valid bulge proposed by \citet{peter2025variability}, this approach introduces, for the first time, a quantitative framework that standardizes the characterization of bulge morphology. Type 1 refers to a bulge that is clearly visible in the electron density profile as a local maximum above the $N_{m, \mathrm{V2}}$. This is detected as an inflection feature in $(N_{s})'$ within the altitude range between $h_{m, \mathrm{BL}}$ and $h_{\mathrm{BL}, 1}$. If one or more sign inversion points are found between these two altitudes, the profile is classified as Type 1. Typically, there are at least two such sign inversion points—one corresponding to a local minimum where the electron density begins to rise again due to the presence of a bulge above V2 peak, and the other marking the actual bulge peak from where the density starts decreasing with altitude. To ensure the presence of a well-defined bulge, we apply an additional criterion that at least five data points should exist between the two inversion points. If no sign inversion is detected in $(N_{s})'$, we examine the profile for Type 2 or Type 3 morphologies. Type 2 morphology refers to a shoulder-like feature above the $N_{m, \mathrm{V2}}$, which is identified when the ratio $m_{\mathrm{BL}, \mathrm{top}} / m_{\mathrm{BL}, \mathrm{bottom}} > 1$, where $m_{\mathrm{BL}, \mathrm{top}}$ is the average value of $(N_{s})'$ in the altitude range between $h_{m, \mathrm{BL}}$ and the altitude where the density is $0.8 \times N_{m, \mathrm{BL}}$ above $h_{m, \mathrm{BL}}$, and $m_{\mathrm{BL}, \mathrm{bottom}}$ is the average of $(N_{s})'$ in the corresponding range below $h_{m, \mathrm{BL}}$. Type 3 represents profiles where no visually distinct bulge is observed in the main electron density profile, but a valid bulge is detected through the residual electron density analysis, and in this case, the ratio $m_{\mathrm{BL}, \mathrm{top}} / m_{\mathrm{BL}, \mathrm{bottom}} < 1$. Profiles that do not satisfy the bulge validity criteria are labeled as uncategorized. The occurrence rates of the different ionospheric morphologies are summarized in Table~\ref{tab:v3_occurrence}.

\begin{table}[ht]
    \centering
    \caption{Summary of occurrence rates of the bulge for different profile categories. Out of all 234 profiles, 29 profiles are uncategorized.}
    \renewcommand{\arraystretch}{1.3} % Adjust row height
    \setlength{\tabcolsep}{10pt} % Adjust column spacing
    \begin{tabular}{|c|c|c|}
        \hline
        \textbf{Category} & \textbf{No. of profiles} & \textbf{occurrence rate (\%)}\\
        \hline
        Type 1 & 26& 11.11\\
        Type 2 & 57& 24.36\\
        Type 3 & 122& 52.14\\
        Uncategorized & 29& 12.39\\
        \hline
    \end{tabular}
    \label{tab:v3_occurrence}
\end{table}

%\textbf{Step - 6: Reanalysis of the selection criteria. 
%%%%%%%%%%%%%%%%%%%%%%%%%%%%%%%%%%%%%%%%%%%%
% Solar activity indices
%%%%%%%%%%%%%%%%%%%%%%%%%%%%%%%%%%%%%%%%%%%%%
\subsection{Classification of solar activity}
\label{Solar activity indices}
 The occurrence of the topside bulge has been previously reported for both high and low solar activity conditions. In this study, the solar activity is classified using the Earth-based $F_{10.7P}$ index, obtained from the National Space Science Data Center (NSSDC) OMNIWeb data set at \url{https://omniweb.gsfc.nasa.gov/form/dx1.html} \citep{mathews1995nssdc}. The $F_{10.7P}$ index represents the mean of the daily solar flux ($F_{10.7}$) and its 81-day running average ($F_{10.7A}$). Since these indices are observed on Earth, they must be processed before being applied to Venus \citep{peter2014dayside, girazian2015characterization}.
 Corrections were applied for the slight variation in the Venus–Sun distance and for differences arising from Earth and Venus observing different hemispheres of the Sun. A “shifted date” was calculated, defined as the date on which Earth was facing the same solar hemisphere that Venus observed on the occultation date. The calculation used $t_{\text{shift}} = t_o \pm \Delta t$ with 
$\Delta t = (27\,\text{days})\left(\tfrac{\text{Earth--Sun--Venus angle}}{360^\circ}\right)$, considering $27~days$ as average solar rotation period, where the $\pm$ sign corresponds to Venus trailing or leading Earth, respectively.
If $|\Delta t| < 7$ days, then the representation of solar activity at Venus $F_{10.7,\text{Venus}}$ is calculated by using
\begin{equation}
F_{10.7P,\mathrm{Venus}}(t_o) =
\left(\frac{1~\mathrm{AU}}{d(t_o)}\right)^2
F_{10.7P}(t_{\text{shift}}).
\label{eq:F107_short}
\end{equation}
If $|\Delta t| >= 7$ days, then a weighted average is computed by using:
\begin{equation}
F_{10.7P,\mathrm{Venus}}(t_o) =
\left(\frac{1~\mathrm{AU}}{d(t_o)}\right)^2
\Big[w_1 F_{10.7P}(t_{\text{shift}}) +
     w_2 F_{10.7P}(t_{\text{shift}} \mp 27~\text{days})\Big],
\label{eq:F107_weighted}
\end{equation}
 Here, $d(t_{o})$ is the Venus-Sun distance in AU during the observation, $w_1 = \left( 0.5 + (13.5 - \Delta t)/13 \right)$, $w_2 = 1 - w_1$, and $F_{10.7P}(X)$ is the $F_{10.7P}$ value measured at 1 AU on day $X$. The thresholds $\Delta t = 7$ and $\Delta t = 20$ correspond to geometrical configurations where Venus and Earth are located on the same side of the Sun, as viewed from Earth. The expression for $w_1$ ensures a proper transition to unshifted values at $\Delta t = 7$ days and $\Delta t = 20$ days. This correction process is imperfect, as it neglects time variations in the solar flux, but it is sufficient for our purposes. The calculated values of $F_{10.7P, Venus}$ are shown in Figure~\ref{Figure_5}.

These solar activity indices are now used to classify the electron density profiles into low, moderate, and high solar activity bins. An electron density profile is categorized as low solar activity if $F_{10.7P,\text{Venus}} \leq 200~sfu$, high solar activity if $F_{10.7P,\text{Venus}} \geq 250~sfu$, and moderate solar activity otherwise $(1~\mathrm{sfu} = 10^{-22}~\mathrm{W\,m^{-2}\, Hz^{-1}})$
.
%%%%%%%%%%%%%%%%%%%%%%%%%%%%%%%%%%%%%%%%%%%%
% Solar Wind data
%%%%%%%%%%%%%%%%%%%%%%%%%%%%%%%%%%%%%%%%%%%%%
\subsection{Data for investigating the solar wind influence on the bulge}
\label{Solar_wind}
\begin{figure}[ht!]
    \centering
    \includegraphics[width=0.95\textwidth]{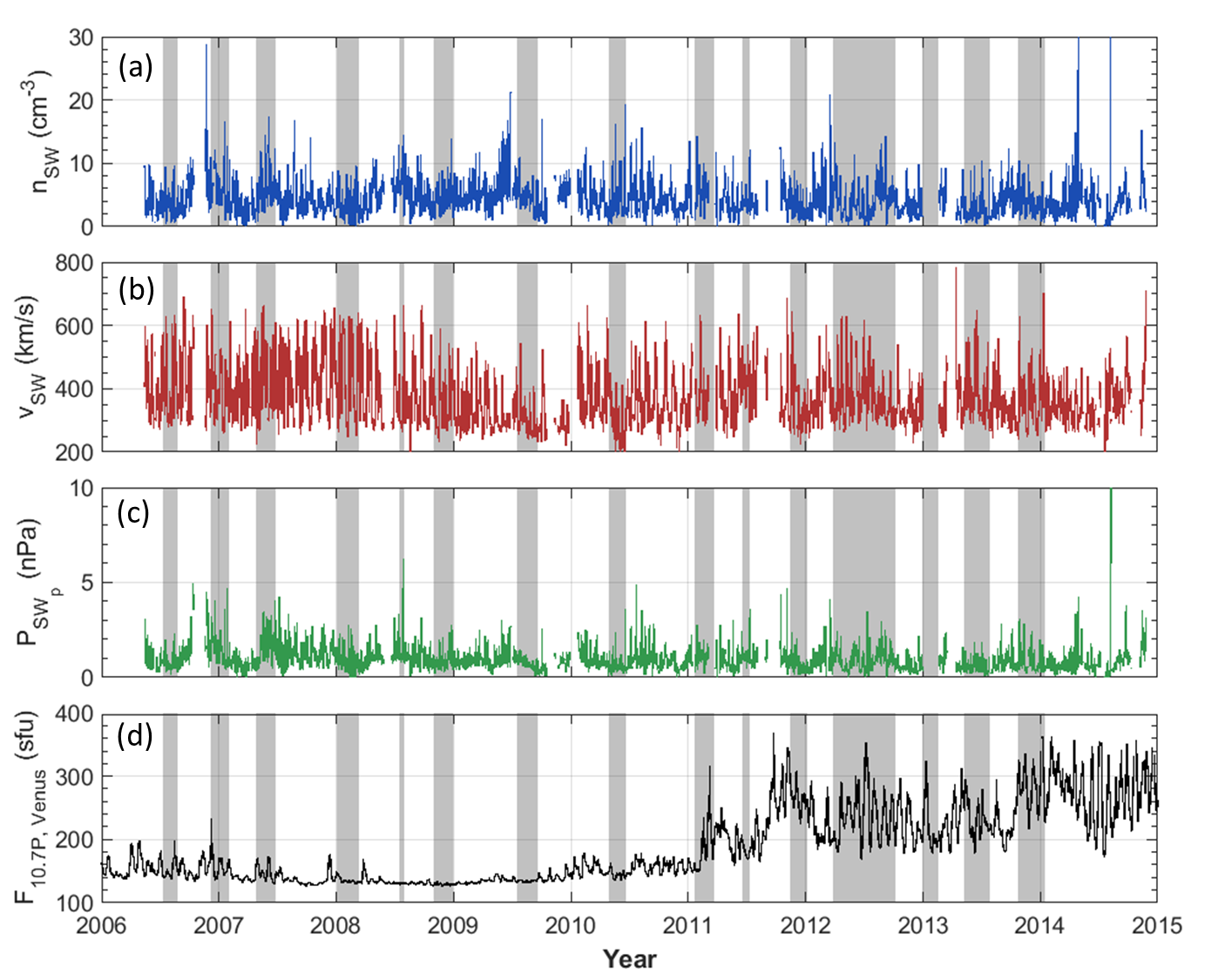}
        \caption{
            Panels (a) and (b) show the solar wind number density ($n_{SW}$) and velocity ($v_{SW}$) at Venus derived from VEX ASPERA-4. 
            Panel (c) shows the solar wind dynamic pressure (eq.~\ref{eq: pristin_SW}). 
            Panel (d) shows the solar activity index $F_{10.7P}$ scaled to Venus. 
            The gray bands mark the occultation seasons (occ~8 was not carried out due to operational limits).
            }
            
    \label{Figure_5}
\end{figure}

In the absence of an intrinsic magnetic field, Venus is directly exposed to the solar wind, which interacts with its topside ionosphere. Variability in the solar wind conditions near Venus can therefore influence the characteristics of the bulge. To calculate the undisturbed solar wind dynamic pressure, measurements are taken from the \textit{ASPERA-4} instrument onboard \textit{Venus Express}. The instrument provides key solar wind parameters, including bulk velocity and density, in the near-Venus environment \citep{barabash2007analyser} (Figure~\ref{Figure_5}). The undisturbed solar wind dynamic pressure at $0.72AU$, is calculated as:
\begin{equation}
    P_{SW_p} = n_{SW} \cdot m_{SW} \cdot v_{SW}^2,
    \label{eq: pristin_SW}
\end{equation}
where $n_{SW}$ is the number density of solar wind protons ($\mathrm{m^{-3}}$), $m_{SW}$ is the proton mass ($\mathrm{kg}$), and $v_{SW}$ is the solar wind velocity ($\mathrm{m/s}$). This parameter is crucial for characterizing the upstream solar wind conditions during radio occultation (RO) events. For each RO event, solar wind data within a $\pm90$-minute window centered at the time when the radio occultation occurs at an altitude of $130~km$ were considered, provided the data are available \citep{peter2025variability}. 

$P_{SW_p}$ is calculated at the time of entry and exit of the \textit{Venus Express} spacecraft from the region where the solar wind is not influenced by the presence of the Venusian ionosphere (i.e., the bow shock region), as shown in Figure~\ref{Figure_5}. If the entry and/or exit data are not available, the point when the spacecraft is nearest to the planet before entering or after exiting the influence of the Venusian ionosphere is considered. In addition, the average of $P_{SW_p}$ for the undisturbed solar wind within the $\pm90$-minute window is taken. RO profiles for which no solar wind data are available are excluded from the analysis.

To account for the angular orientation of the solar wind relative to the planetary ionosphere, a geometric correction is applied using the flow solar zenith angle ($SZA_F$). The adjusted dynamic pressure at the occultation point is expressed as:
\begin{equation}
    P_{SW} = P_{SW_p} \cdot \cos^2(SZA_F),
\end{equation}
where $SZA_F$ denotes the angle between the upstream solar wind flow direction and the surface normal at the location of radio occultation in the aberration angle corrected Venus Solar Orbital coordinate system  (VSO$'$), which aligns with the actual solar wind velocity vector rather than the Sun–Venus line. The correction accounts for the aberration induced by Venus's orbital motion, quantified by the aberration angle:
\begin{equation}
    \Delta = \arctan\left(\frac{v_O}{v_{SW}}\right),
\end{equation}
where $v_O$ is the orbital velocity of Venus. A nominal value of $\Delta = 5^\circ$ is adopted, following \citet{signoles2023influence}.

The transformation from the standard  Venus Solar Orbital (VSO) frame to VSO$'$ is performed via a rotation about the $+Z$-axis by the angle $\Delta$. In the VSO coordinate frame, the $+X$-axis points from the planet center to the Sun, the $+Y$-axis points in the opposite direction of the orbital velocity, and the $+Z$-axis completes the orthogonal right-handed coordinate system. In the corrected frame (VSO$'$), the $+X$-axis is aligned with the solar wind velocity vector rather than the Sun–Venus line, to account for the aberration due to the orbital motion of Venus. This correction ensures a more accurate representation of the solar wind interaction geometry by aligning the frame with the actual upstream solar wind direction.

%%%%%%%%%%%%%%%%%%%%%%%%%%%%%%%%%%%%%%%%%%%%
%
% Results (or Analysis of V3)
%
%%%%%%%%%%%%%%%%%%%%%%%%%%%%%%%%%%%%%%%%%%%%
\section{Results}
\label{Results}
The Venusian ionospheric bulge has been examined under varying solar activity conditions and across the entire range of solar zenith angles. Table~\ref{tab:profiles_table} summarizes the classification of profile morphologies, together with the corresponding bulge altitudes and peak electron densities.
%%%%%%%%%%%%%%%%%%%%%%%%%%%%%%%%%%%%%%%%%%%
% Morphology of bulge
%%%%%%%%%%%%%%%%%%%%%%%%%%%%%%%%%%%%%%%%%%%
\subsection{Spatio-temporal variability in the morphology of the bulge}  
For investigating the bulge behavior, we categorized the final $234$ profiles into three morphological types—Type 1, Type 2, and Type 3—through an automated classification routine. As discussed in Section~\ref{Electron density}, Type~1 profiles exhibit a distinct local maximum above the V2 layer, where the bulge is clearly evident. Type~2 profiles exhibit a shoulder-like structure at the expected bulge altitude. Type~3 profiles show only subtle variation in the electron density. After fitting, if the residuals reveal a distinct layer-like structure, the profile is classified as Type 3; otherwise, it remains uncategorized. These morphological types qualitatively capture the extent to which the physical mechanisms driving charge accumulation dominate: Type 1 corresponds to strong dominance, while Type 2 and Type 3 reflect progressively weaker effects.
\begin{table}[ht]
    \centering
    \caption{Summary of all 234 VeRa profiles.}
    \resizebox{\textwidth}{!}{%
    \begin{tabular}{|l|l|c|c|c|c|}
        \hline
        \textbf{Solar activity} & \textbf{Morphology} & \textbf{No. of Profiles} & \textbf{Occurrence (\%)} & \textbf{Mean $h_{m, BL}$ ($km$)} & \textbf{Mean $N_{m, BL}$ ($m^{-3}$)} \\
        \hline
        \multirow{3}{*}{\textbf{Low}}          
            & Type 1        & $22$  & $16$     & $179.0 \pm 9.6$    & $6.0 \times 10^{10} \pm 1.9 \times 10^{10}$ \\
            & Type 2        & $37$  & $26$     & $171.6 \pm 8.4$    & $7.5 \times 10^{10} \pm 7.6 \times 10^{9}$ \\
            & Type 3        & $58$  & $42$     & $173.0 \pm 8.7$    & $7.4 \times 10^{10} \pm 1.4 \times 10^{10}$ \\
            & Uncategorized & $23$  & $16$     & -                  & - \\
        \hline
        \multirow{3}{*}{\textbf{Medium}} 
            & Type 1        & $3$   & $7$       &$180.5 \pm 14.2$    & $8.5 \times 10^{10} \pm 4.6 \times 10^{10}$ \\
            & Type 2        & $11$  & $28$      & $166.1 \pm 3.6$    & $5.2 \times 10^{10} \pm 2.0 \times 10^{10}$ \\
            & Type 3        & $26$  & $65$      & $167.4 \pm 3.8$    & $4.7 \times 10^{10} \pm 1.7 \times 10^{10}$ \\
            & Uncategorized & $0$   & $0$       & -                  & - \\
        \hline
        \multirow{3}{*}{\textbf{High}}
            & Type 1        & $1$   & $2$       & $189.2 \pm 1.6$    & $1.1 \times 10^{11} \pm 3.0 \times 10^9$ \\
            & Type 2        & $9$   & $17$      & $164.0 \pm 3.4$    & $7.2 \times 10^{10} \pm 1.6 \times 10^{10}$ \\
            & Type 3        & $38$  & $70$      & $168.5 \pm 6.4$    & $8.0 \times 10^{10} \pm 2.5 \times 10^{10}$ \\
            & Uncategorized & $6$   & $11$      & -                  & - \\
        \hline
        %\textbf{All\_Types} & All & 113 & $167.1 \pm 4.0$ & $6.1 \times 10^4 \pm 2.4 \times 10^4$ \\  
        %\hline
    \end{tabular}%
    }
    %\raggedright
    %{\footnotesize * The error in $h_{m,\mathrm{BL}}$ and $N_{m,\mathrm{BL}}$ is the values within $1\sigma$. For the single data points, the error in $N_{m,\mathrm{BL}}$ is assigned to that of the profile, while for $h_{m,\mathrm{BL}}$, the error is calculated by the method described in \citep{girazian2015characterization}.}
    \label{tab:profiles_table}
\end{table}

\begin{figure}[ht]
    \centering
    \includegraphics[width=1\linewidth]{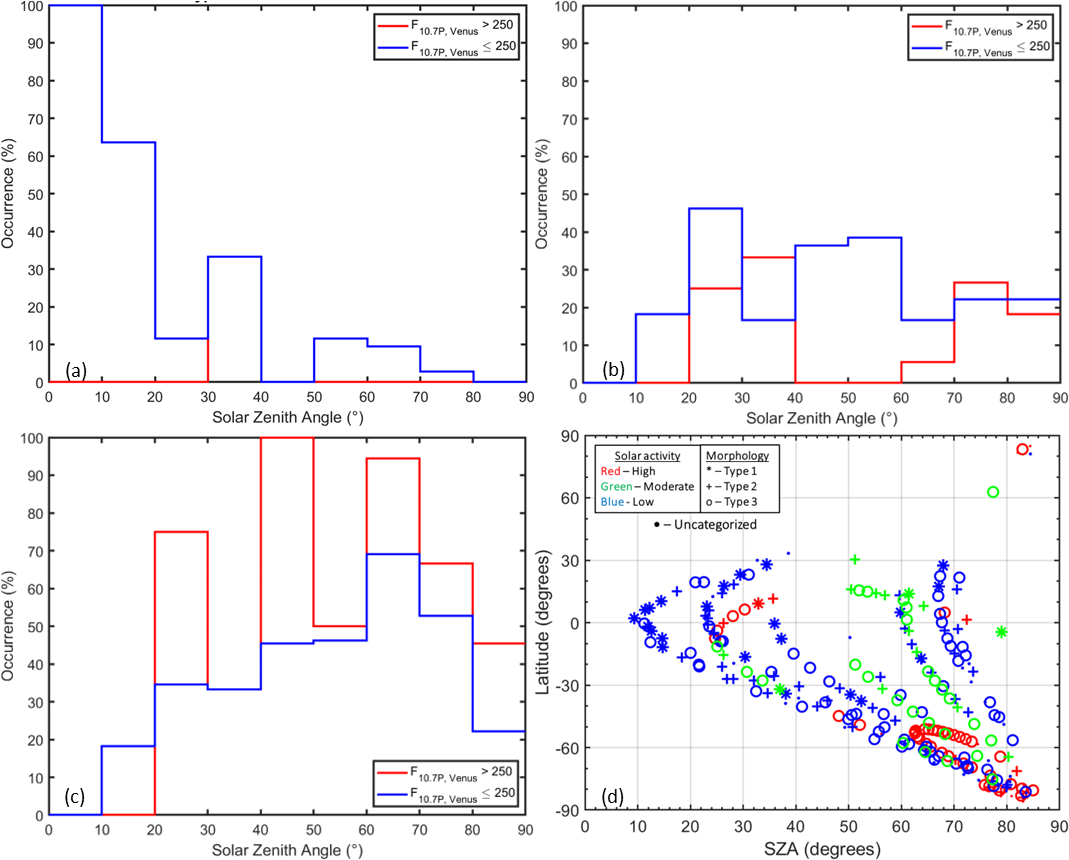}
    \caption{(a) Occurrence rate of Type 1 profiles as a function of $SZA$. The highest occurrence is seen at low $SZA$, with a clear decrease in occurrence rate as $SZA$ increases. 
    (b) Occurrence rate of Type 2 profiles as a function of $SZA$, showing a reduced occurrence at higher $SZA$. (c) Occurrence rate of Type 3 profiles with $SZA$, which also decreases at higher $SZA$.} 
    (d) Distribution of bulge occurrences as a function of latitude and corresponding Solar Zenith Angle ($SZA$). Type 1 bulges are observed predominantly at low latitudes, mostly within $\pm 40^\circ$.
    \label{Figure_6}
\end{figure}

The bulge morphology exhibits a clear dependence on solar zenith angle ($SZA$), transitioning from Type 1 at low $SZAs$ to Type 3 at higher $SZAs$ (Figure~\ref{Figure_6}). The occurrence rate of Type 1 morphology is high at low $SZAs$ but declines with increasing $SZA$. Type 1 morphology is not observed when the $SZA$ exceeds $80^\circ$ in the dayside ionosphere, indicating that the processes responsible for bulge formation are more effective in the subsolar region. This finding is consistent with earlier observations from the Pioneer Venus Orbiter (\textit{PVO}), where \citet{woo1991magnetization, ivanov1979daytime} reported a prominent bulge in electron density profiles near the subsolar region, which diminishes to a small kink at higher $SZAs$. 

Because Venus has a small axial tilt of only $2.7^\circ$, the solar zenith angles ($SZA$) at higher latitudes remain consistently large throughout the year. At a given latitude, the minimum $SZA$ is approximately equal to the latitude itself; hence, low $SZAs$ cannot occur at high latitudes. Analysis of the VeRa dataset reveals that Type 1 morphology is largely restricted to low latitudes ($\pm 40^\circ$) (Figure \ref{Figure_6}). Notably, a substantial fraction of the $234$ analyzed profiles originates from the southern hemisphere. Nevertheless, despite this large dataset and the broad coverage of solar activity and $SZA$ conditions, no instances of Type 1 morphology were detected beyond ($\pm 40^\circ$) latitude. This finding - especially the complete absence of Type 1 morphology in the predominantly observed southern hemisphere - strongly indicates that this bulge morphology is confined to low-latitude regions (Table~\ref{tab:Type1bulge_data_table}).
  
\begin{figure}[ht]
    \centering
    \includegraphics[width=0.8\textwidth]{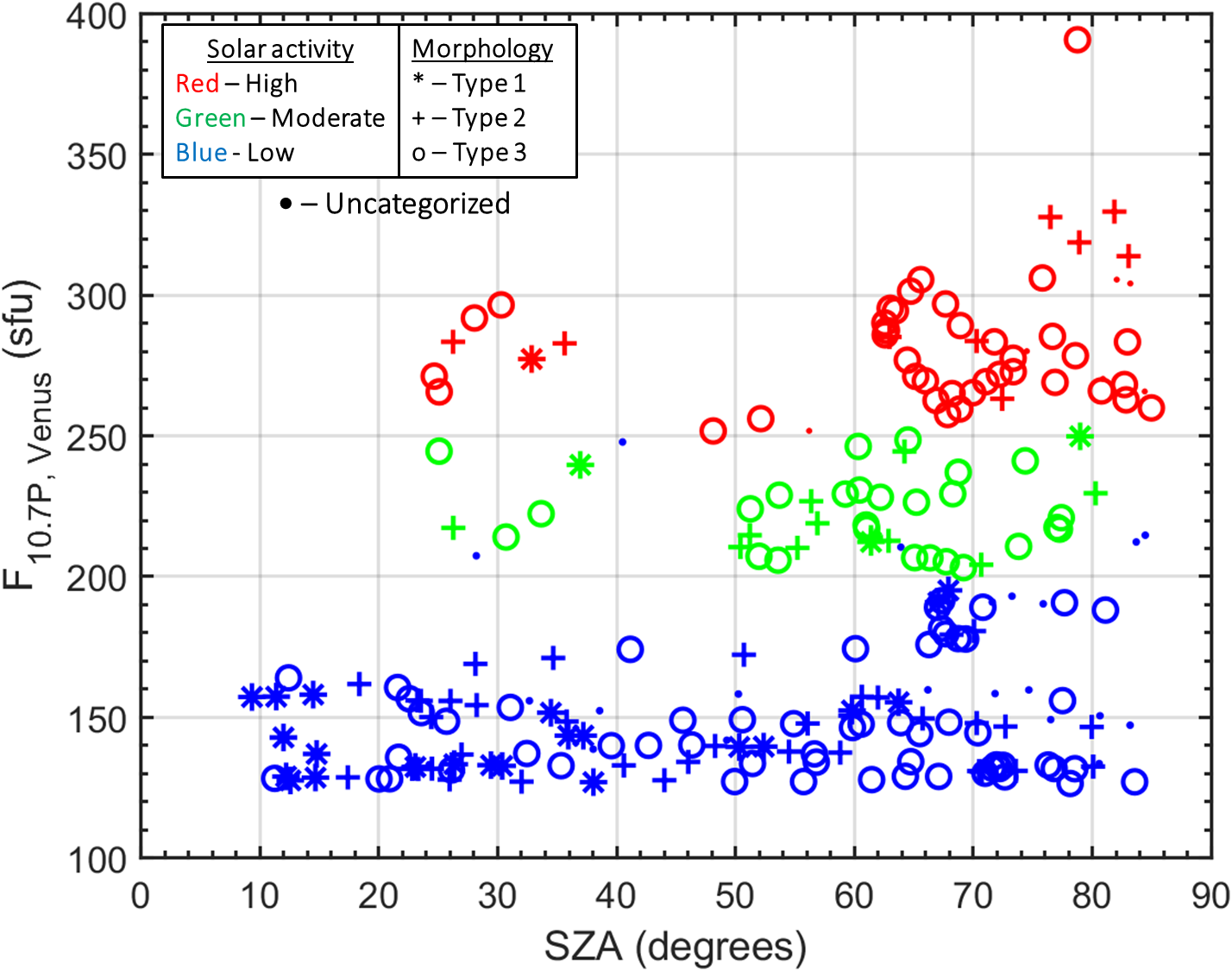}
    \caption{Observations of the Venusian ionosphere across varying solar activity levels and solar zenith angles.}
    \label{Figure_7}
\end{figure}
 
Only one profile with Type 1 morphology was observed during high solar activity, and three profiles during moderate solar activity, while the majority occurred during low solar activity (Figure~\ref{Figure_7}). This distribution suggests that the dominant physical processes responsible for the bulge become less effective as solar activity increases. This observed trend may reflect modifications within the Venusian ionosphere and/or variations in the space environment influenced by solar activity.

\begin{table}[ht]
    \centering
    \caption{All the Type 1 profiles}
    \resizebox{\textwidth}{!}{%
    \begin{tabular}{|c|c|c|c|c|c|c|c|c|}
        \hline
        \textbf{Year} & \textbf{DOY} & \textbf{I/E} & \textbf{$F_{10.7P,\ Venus}$ (sfu)} & \textbf{LAT ($^\circ$)} & \textbf{LON ($^\circ$)} & \textbf{$SZA$ ($^\circ$)} & \textbf{$h_{m,\mathrm{BL}}$ ($km$)} & \textbf{$N_{m,\mathrm{BL}}$ ($\times10^{10}$ m$^{-3}$)} \\
        \hline
        2007 & 020 & E & 155 & -17.28 & 282.85 & 63.71 & 173.7 & 5.01 \\
        2007 & 026 & E & 152 & 4.85 & 299.64 & 59.79 & 174.5 & 4.58 \\
        2007 & 173 & I & 137 & -11.7 & 324.88 & 14.78 & 179.9 & 7.42 \\
        2007 & 176 & I & 143 & -1.85 & 331.56 & 11.99 & 176.8 & 4.17 \\
        2008 & 011 & E & 144 & -0.54 & 203.92 & 35.94 & 199.6 & 3.26 \\
        2008 & 013 & E & 144 & -7.87 & 209.24 & 37.2 & 185.4 & 4.16 \\
        2008 & 021 & E & 139 & -34.62 & 230.64 & 50.31 & 194.6 & 5.01 \\
        2008 & 022 & E & 140 & -37.75 & 233.33 & 52.37 & 173.8 & 6.43 \\
        2008 & 352 & E & 127 & -34.06 & 225.14 & 38.05 & 193.9 & 3.05 \\
        2008 & 360 & E & 129 & -7.62 & 247.13 & 14.68 & 184.9 & 5.26 \\
        2008 & 361 & E & 128 & -4.05 & 249.81 & 12.57 & 172.8 & 7.47 \\
        2008 & 364 & E & 129 & 7.15 & 257.81 & 12.21 & 190.6 & 3.66 \\
        2009 & 197 & I & 133 & 23.02 & 118.68 & 29.44 & 178.3 & 6.9 \\
        2009 & 198 & I & 133 & 17.55 & 121.43 & 26.35 & 176.5 & 5.83 \\
        2009 & 200 & I & 132 & 7.83 & 126.88 & 23.08 & 164.8 & 7.86 \\
        2009 & 206 & I & 133 & -16.45 & 143.07 & 30.38 & 177.2 & 6.50 \\
        2010 & 191 & I & 152 & 27.96 & 189.03 & 34.47 & 170.2 & 6.34 \\
        2011 & 021 & E & 158 & 10.39 & 43.01 & 14.49 & 183.6 & 6.02 \\
        2011 & 022 & E & 157 & 6.17 & 45.41 & 11.38 & 179.2 & 6.55 \\
        2011 & 023 & E & 157 & 2.1 & 47.82 & 9.34 & 177.9 & 5.92 \\
        2011 & 165 & I & 195 & 27.61 & 67.1 & 67.90 & 165.4 & 10.8 \\
        2011 & 167 & I & 191 & 17.43 & 72.49 & 67.06 & 164.4 & 9.43 \\
        2012 & 163 & I & 250 & -4.43 & 251.98 & 78.97 & 189.6 & 3.31 \\
        2012 & 190 & I & 277 & 9.15 & 288.03 & 32.86 & 189.2 & 10.6 \\
        2012 & 212 & I & 240 & -32.03 & 336.65 & 36.95 & 187.7 & 12.0 \\
        2013 & 365 & E & 212 & 13.87 & 16.23 & 61.39 & 164.2 & 10.3 \\
        \hline
    \end{tabular}%
    }
    \raggedright
    {\footnotesize
    \begin{enumerate}
        \item DOY stands for Day Of Year.
        \item \textbf{I} and \textbf{E} denote \textbf{Ingress} and \textbf{Egress}, respectively.
        \item sfu refers to the Solar Flux Unit, where $1~\mathrm{sfu} = 10^{-22}~\mathrm{W\,m^{-2}\,Hz^{-1}}$.
    \end{enumerate}
    }
    \label{tab:Type1bulge_data_table}
\end{table}

%%%%%%%%%%%%%%%%%%%%%%%%%%%%%%%%%%%%%%%%%%%%%%%%%%%%%
% SZA variation of hmV3
%%%%%%%%%%%%%%%%%%%%%%%%%%%%%%%%%%%%%%%%%%%%%%%%%%%%%
\subsection{Variation of peak altitude ($h_{m, BL}$) and peak density ($N_{m, BL}$) of bulge with $SZA$}
\begin{figure}[ht]
    \centering
    \includegraphics[width=0.5\linewidth]{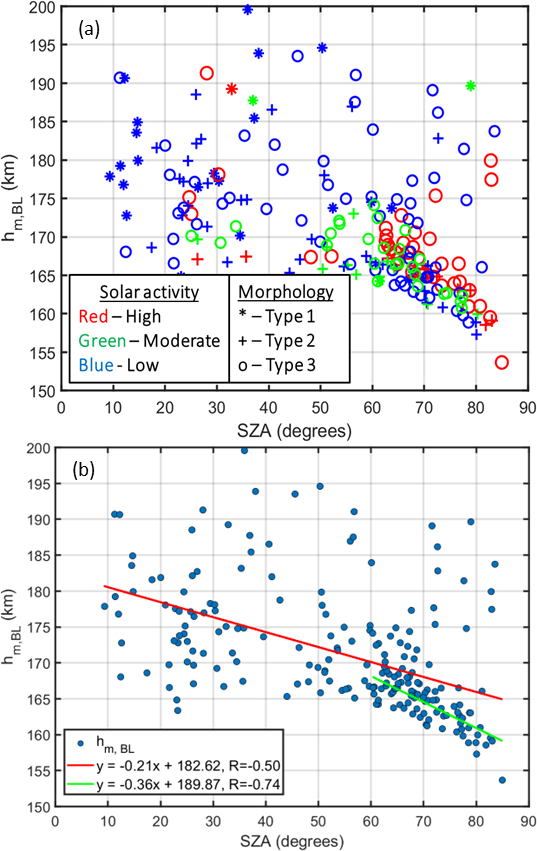}
    \caption{
        The figure panels illustrate the variability of $h_{m, BL}$ as a function of Solar Zenith Angle ($SZA$). 
        The panels display the $h_{m, BL}$ for Types 1, 2, and 3 profiles (top) and corresponding fit parameters (bottom). 
        The red line represents a linear fit, with the R-value of the fit shown. The green line represents a subset of the data for $SZA\geq60^\circ$ and $h_{m, BL} < 170~km$, showing a sharp decrease in altitude of the bulge.
    }
    \label{Figure_8}
\end{figure}

Figure~\ref{Figure_8} shows the dependence of the bulge altitude on solar zenith angle ($SZA$). We observe that the peak altitude of the bulge ($h_{m, BL}$) decreases with increasing $SZA$. This behavior contrasts with the $h_{m, V2}$ and $h_{m, V1}$ layers, which remain relatively constant for $SZA$ values below ($80^\circ$) \citep{girazian2015characterization,gerard2017aeronomy}. The dependence of electron density peak altitudes in any atmosphere is mainly controlled by (i) the solar radiation path length traversed in the atmosphere and (ii) the background neutral density. Venus’s slow rotation results in significant cooling of the atmosphere near the terminator, causing a collapse of the neutral atmosphere at higher $SZAs$. As a result, the longer solar radiation path is counterbalanced by lower atmospheric density at the same altitude, leading to the observed insensitivity of $h_{m, V2}$ and  $h_{m, V1}$ to $SZA$ variations. At higher altitudes, however, the variation in neutral density with $SZA$ becomes more pronounced \citep{hedin1983global}. In particular, neutral density in the thermosphere decreases significantly before $8~\mathrm{hrs}$ and after $16~\mathrm{hrs}$ local time, corresponding to $SZA \gtrsim 60^\circ$. This pronounced drop in density at higher $SZA$ values exerts a stronger influence on the bulge altitude. Our analysis shows a strong negative correlation between the peak altitude of the bulge layer ($h_{m, BL}$) and $SZA$ for the subset of data where $SZA \geq 60^\circ$ (considering $h_{m, BL}\leq 170~km$, Figure~\ref{Figure_8}). This strong correlation suggests that the peak altitude of the bulge is predominantly controlled by the significant variation in thermospheric neutral density. This dependence on neutral density at higher altitudes also suggests that the bulge may be merging with the V2 layer at high $SZAs$.

\begin{figure}[ht]
    \centering
    \includegraphics[width=0.5\linewidth]{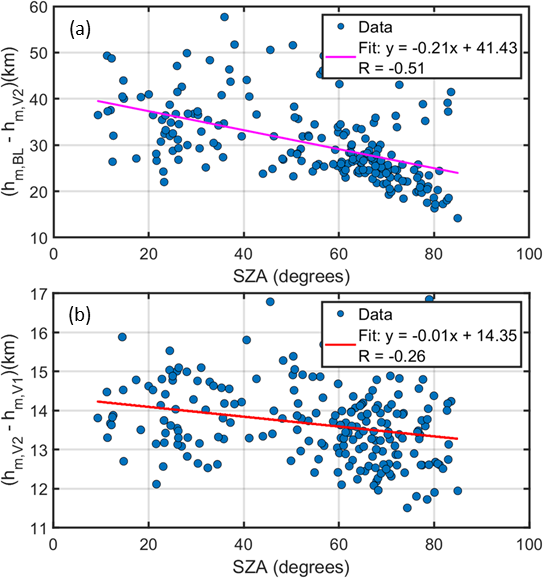}
    \caption{
    In this figure, the top panel shows the variation of $h_{m, BL} - h_{m, V2}$, which exhibits a moderate correlation with $SZA$ due to the significant dependence of $h_{m, BL}$ on $SZA$. 
    In contrast, the bottom panel shows a weak correlation of $h_{m, V2} - h_{m, V1}$ with $SZA$, indicating that these layers are relatively invariant to $SZA$. }
    \label{Figure_9}
\end{figure}

We calculated the difference between $h_{m,BL}$ and $h_{m,V2}$ to further investigate the observed trend (Figure~\ref{Figure_9}). The separation between the two peak altitudes decreases with increasing $SZA$, consistent with the result that $h_{m, V2}$ remains nearly constant while $h_{m, BL}$ shifts downward with increasing $SZA$. A similar behavior is reported for the Martian ionosphere, where the separation between $h_{m, M3}$ and $h_{m, M2}$ also decreases with increasing $SZA$ \citep{mayyasi2018sporadic}. However, the underlying mechanism is different: at Mars, the decrease in separation arises because $h_{m, M2}$ increases with $SZA$, while $h_{m, M3}$ remains relatively insensitive. In contrast, at Venus, it is the bulge altitude ($h_{m, BL}$) that responds strongly to thermospheric density variations, while $h_{m, V2}$ remains largely unaffected by $SZA$. This highlights that although both planets exhibit a reduction in layer separation with $SZA$, the physical drivers governing these trends are distinct.

\begin{figure}
    \centering
    \includegraphics[width=1\linewidth]{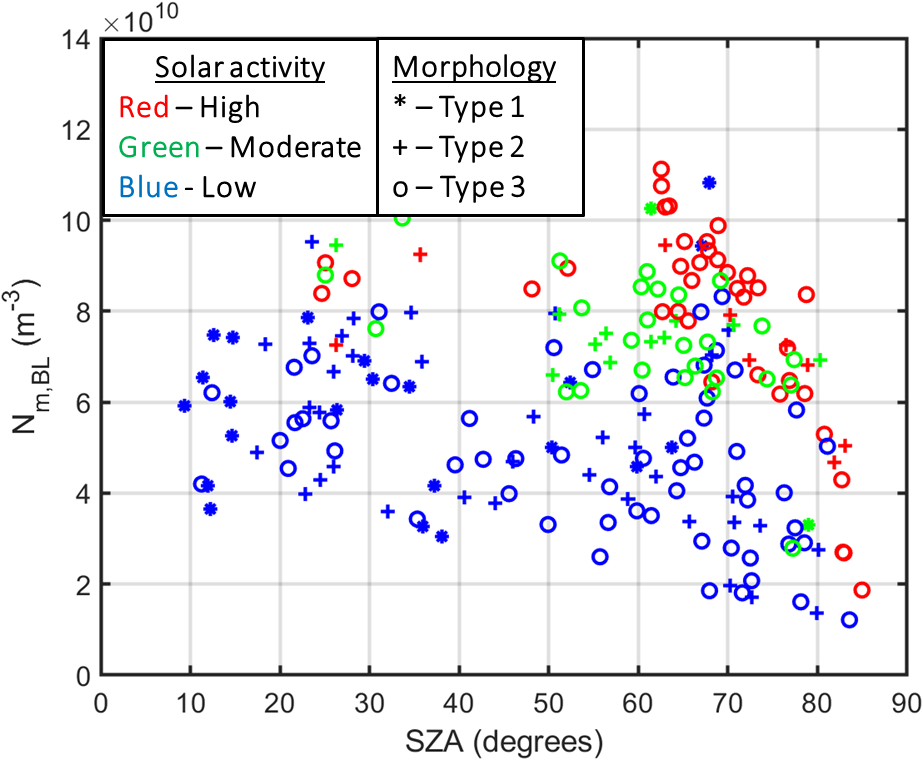}
    \caption{
    Variation of the peak density of the bulge $N_{m, BL}$ with solar zenith angle.
    $N_{m,\mathrm{BL}}$ values are generally higher during periods of high solar activity. }
    \label{Figure_10}
\end{figure}

%%%%%%%%%%%%%%%%%%%%%%%%%%%%%%%%%%%%%%%%%%%%%%%%
% SZA variation of nmV3
%%%%%%%%%%%%%%%%%%%%%%%%%%%%%%%%%%%%%%%%%%%%%%%%
We now turn to the variability of the bulge’s peak electron density ($N_{m, BL}$). For the main ionospheric layers (V1 and V2), it is well established \citep{ivanov1979daytime, breus1985properties, peter2014dayside, girazian2015characterization, gerard2017aeronomy} that the peak electron density ($N_{m,V1}, N_{m,V2}$) follows the Chapman prediction, decreasing with increasing solar zenith angle ($SZA$) as the available photon flux for ionization diminishes. In contrast, the bulge exhibits only a weak correlation between its peak density and $SZA$ (Figure \ref{Figure_10}). This variability indicates that the processes responsible for the bulge are irregular compared to the well-understood photochemical control of the V1 and V2 layers. This will be discussed later in Section~\ref{Discussion}.

\begin{figure}[ht]
    \centering
    \includegraphics[width=0.5\linewidth]{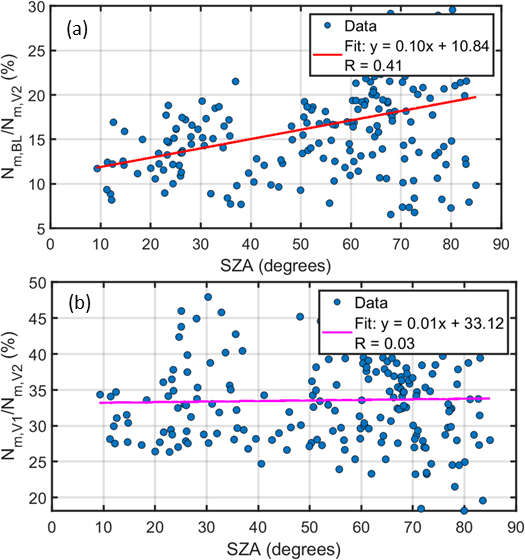}
    \caption{Upper panel: Variation of the ratio $N_{m,\mathrm{BL}}/N_{m,\mathrm{V2}}$ with Solar Zenith Angle ($SZA$). A moderate positive correlation is seen, as the ratio increases with larger $SZA$. This is because $N_{m,\mathrm{V2}}$ decreases approximately as $\cos^{0.5}(\mathrm{SZA})$ following Chapman layer behavior, while $N_{m,\mathrm{BL}}$ remains nearly constant or only weakly related to $SZA$. As a result, the ratio increases with $SZA$. Lower panel: The ratio $N_{m,\mathrm{V1}}/N_{m,\mathrm{V2}}$ shows little dependence on $SZA$, since both $N_{m,\mathrm{V1}}$ and $N_{m,\mathrm{V2}}$ vary in a similar way with $SZA$.}
    \label{Figure_11}
\end{figure}

A more detailed analysis of the variability of the ratios $N_{m, BL}/N_{m,V2}$ and $N_{m, V1}/N_{m,V2}$ with SZA reveals distinct patterns. While $N_{m, V1}/N_{m,V2}$ remains nearly constant with $SZAs$, $N_{m, BL}/N_{m,V2}$ exhibits a moderate positive correlation. The constancy of $N_{m, V1}/N_{m, V2}$ indicates that the processes controlling the V1 and V2 layers are relatively stable across different $SZAs$ (Figure~\ref{Figure_11}). This arises because the peak densities of both the V2 and V1 layers follow the Chapman behavior,  as discussed earlier, and therefore vary similarly with $SZA$ \citep{girazian2015characterization, gerard2017aeronomy}. The relatively large scatter in $N_{m, V1}/N_{m, V2}$ is attributed to the strong variability of solar soft X-ray flux, and consequently in $N_{m, V1}$, compared to the more stable solar EUV radiation. This variability is further amplified during periods of high solar activity.

However, a positive trend observed in the ratio $N_{m, BL}/N_{m, V2}$ with $SZA$ (Figure~\ref{Figure_11}) is primarily governed by the strong dependence of $N_{m, V2}$ on $SZA$: as $N_{m, V2}$ decreases consistently with increasing SZA, the ratio rises even though $N_{m, BL}$ remains relatively constant across all $SZAs$. This non-Chapman behavior of $N_{m, BL}$ underscores the role of additional, variable sources or mechanisms influencing the bulge, which are distinct from the processes controlling the V2 layer, as will be discussed in Section~\ref{Discussion}.
% However, the lower $N_{m, BL}$ values at higher $SZAs$ can also be attributed to the lower neutral density near the terminator because of the low temperature, resulting in a smaller neutral scale height.

%It is noteworthy that the slight increasing trend in$N_{m, V1}/N_{m,V2}$ may be attributed to:  
%a) The predominance of data collected during high solar activity at higher SZAs, and  
%b) The $N_{m, V1}$ density’s greater sensitivity to solar activity due to the hardening of the solar spectrum \citep{girazian2015characterization}, which leads to an increase in the $N_{m, V1}/N_{m,V2}$ ratio.  

%%%%%%%%%%%%%%%%%%%%%%%%%%%%%%%%%%%%%%%%%%%%
% Solar activity variation of hmBL
%%%%%%%%%%%%%%%%%%%%%%%%%%%%%%%%%%%%%%%%%%%%
\subsection{Influence of solar activity on peak altitude ($h_{m, \mathrm{BL}}$) and peak density ($N_{m, \mathrm{BL}}$) of the bulge}  
\begin{figure}[ht]
    \centering
    \includegraphics[width=0.5\linewidth]{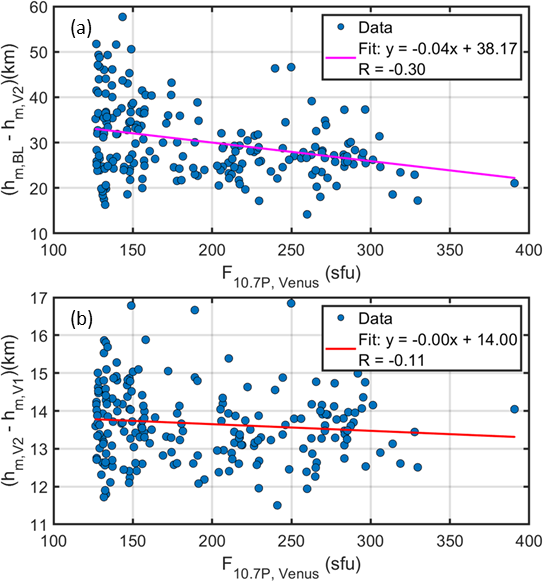}
    \caption{The top panel illustrates the variation of $h_{m, BL} - h_{m, V2}$, while the bottom panel depicts $h_{m, V2} - h_{m, V1}$ with Solar activity. Both altitude differences show no correlation with the $F_{10.7P, Venus}$.}% Both plots reveal a weak correlation with Solar activity. However, it is important to note the limited data available during periods of high Solar activity at lower $SZA$, as V3 is more readily identified at lower $SZA$.     Consequently, fewer observations of the bulge were made during these periods.
    \label{Figure_12}
\end{figure}

The peak altitude of the bulge ($h_{m, \mathrm{BL}}$) shows no clear correlation with solar activity. \citet{girazian2015characterization} reported no significant variation of the V1 and V2 peak altitudes with solar activity. Their analysis of VeRa observations showed that the average altitudes of both layers remained nearly constant across high, medium, and low solar activity conditions. Similarly, \citet{gerard2017aeronomy} found $h_{m, V2}$ to be stable at about $140.7~km$ for $SZA < 80^\circ$. In contrast, \citet{brace1991structure} noted an approximate $5~km$ decrease in the V2 peak altitude from solar maximum to solar minimum in unpublished \textit{PVO} data, but without a clear explanation. For additional insight, we compared the altitude differences $h_{m, BL} - h_{m, V2}$ and $h_{m, V2} - h_{m, V1}$. Both differences also show only weak dependence on solar activity (Figure~\ref{Figure_12}).  

\begin{figure}[ht]
    \centering
    \includegraphics[width=0.5\linewidth]{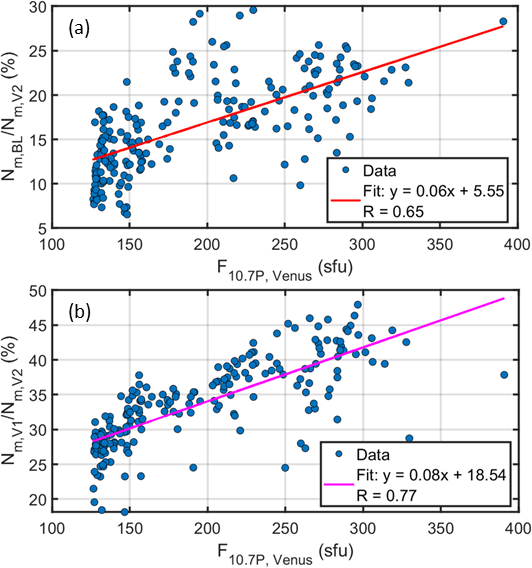}
    \caption{The ratios $N_{m, BL}/N_{m, V2}$ and $N_{m, V1}/N_{m, V2}$ are compared against the solar activity index $F_{10.7P, Venus}$. The $N_{m, BL}/N_{m, V2}$ ratio shows an increase with rising solar activity, potentially linked to the expansion of the neutral thermosphere, a steeper electron density gradient above the V2 layer \citep{gerard2017aeronomy}, and changes in neutral scale height and composition \citep{hensley2020dependence}. In contrast, the $N_{m, V1}/N_{m, V2}$ ratio also increases with solar activity, primarily due to the hardening of the solar spectrum in the soft X-ray range.}
    \label{Figure_13}
\end{figure}

%%%%%%%%%%%%%%%%%%%%%%%%%%%%%%%%%%%%%%%%%%%%
% Solar activity variation of NmBL
%%%%%%%%%%%%%%%%%%%%%%%%%%%%%%%%%%%%%%%%%%%%
In Figure \ref{Figure_10}, we can observe that the electron density at the time of high solar activity is higher across all $SZA$. To analyze the effect of solar activity on the peak density of the bulge ($N_{m, BL}$), we examine the ratios in the same manner as in the preceding section. As illustrated in Figure \ref{Figure_13}, the $N_{m, BL}/N_{m, V2}$ ratio increases with solar activity, a trend that can be explained by the expansion of the neutral thermosphere, as elaborated in the following discussion.

With increasing solar activity, the electron densities of the V2 and V1 layers rise in response to the enhanced solar flux \citep{mendillo2020ionosphere}. The topside ionosphere of Venus, between ($140 - 180~km$), also shows higher electron density at all altitudes under such conditions \citep{gerard2017aeronomy, hensley2020dependence}. In particular, \citet{gerard2017aeronomy} reported that elevated solar activity produced an increased electron density gradient above the V2 layer. the variation in the electron density above the V2 layer is higher compared to $N_{m, V2}$ and increases with the increasing altitude \citep{hensley2020dependence}. This effect arises from thermal expansion driven by enhanced EUV flux, which modifies the neutral scale height and alters atmospheric composition \citep{hensley2020dependence}. Consequently, the increase in electron density above the V2 layer during high solar activity is likely to contribute to the observed value of  $N_{m,BL}$, and therefore the ratio $N_{m, BL}/N_{m, V2}$ increases with the solar activity.

Conversely, the $N_{m, V1}/N_{m, V2}$ ratio exhibits a more straightforward relationship with solar activity. The increase in this ratio during solar maxima can be directly linked to the hardening of the solar spectrum, particularly in the soft X-ray range. However, in the topside atmosphere, it is the variation in the background neutral density with the solar activity that is responsible for the higher electron density at the topside ionosphere. This distinct behavior underscores the varying mechanisms influencing electron density profiles in different ionospheric layers under different solar conditions. 

Although a definitive explanation remains elusive, existing research, as discussed above, offers valuable insights into the effect of solar activity on $N_{m, BL}$. More research is needed to fully elucidate the processes driving the observed trend.

\subsection{Solar wind influence on the bulge}
\begin{figure}
    \centering
    \includegraphics[width=1\linewidth]{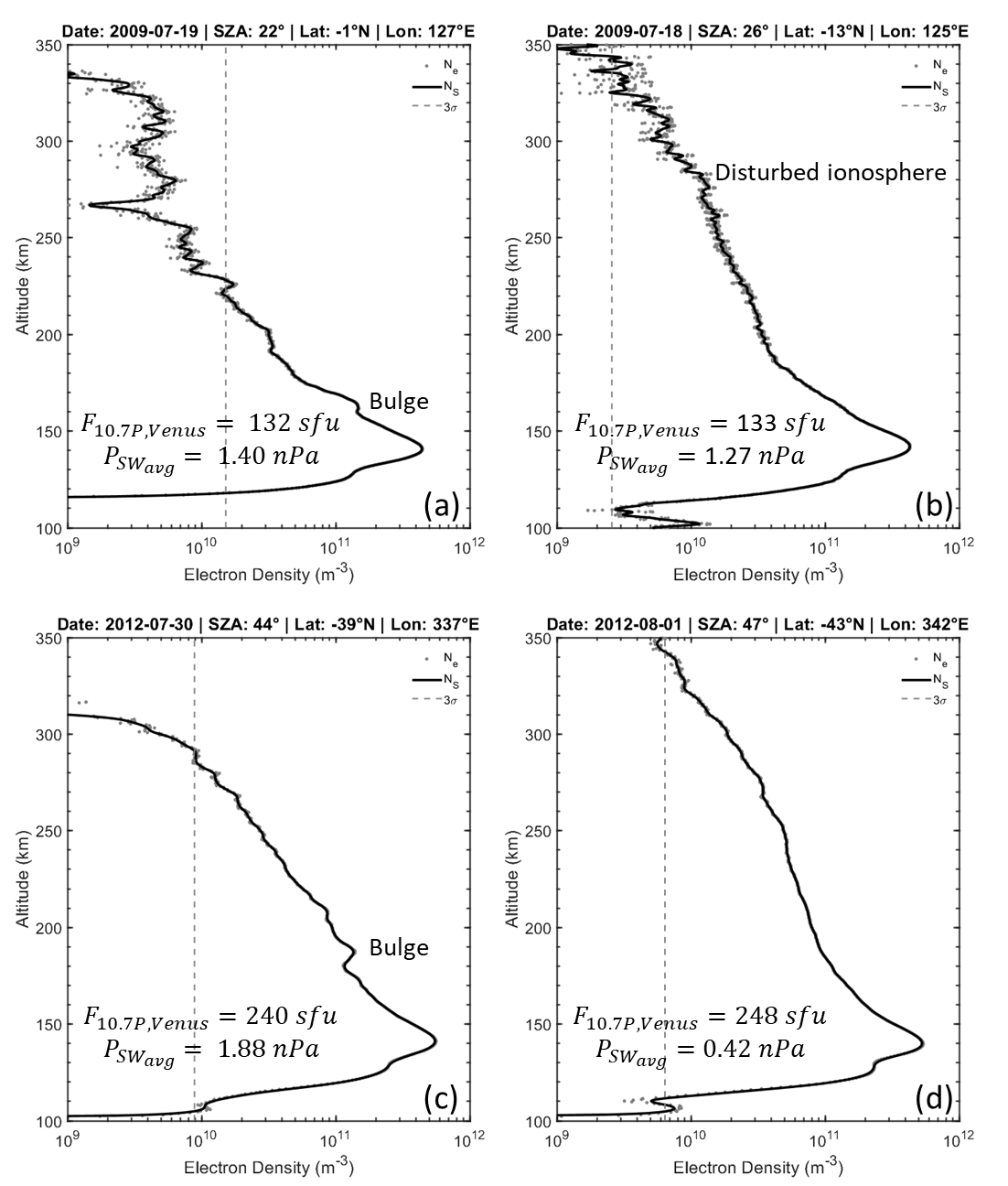}
    \caption{Influence of solar wind dynamic pressure on the Venusian ionosphere and bulge morphology. 
    (a) and (b) show a Type~1 bulge profile and a disturbed topside ionosphere respectively, under similar solar wind dynamic pressure and solar activity conditions. 
    (c) illustrates a Type~1 profile at high solar wind dynamic pressure, while (d) shows the absence of a bulge at low solar wind dynamic pressure, both corresponding to similar solar activity levels. The vertical dotted gray line indicates the $3\sigma$ noise level in the profile.}
    \label{Figure_14}
\end{figure}

Unlike Earth, Venus does not generate an intrinsic magnetic field, allowing the solar wind to interact directly with the Venusian ionosphere. The variability of the Venusian topside ionosphere in response to solar wind dynamic pressure has been examined in detail by \citet{peter2025variability} and references therein. In this work, the solar wind dynamic pressure at Venus is analyzed for its effect on the occurrence and morphology of the bulge (Figure~\ref{Figure_5}). The solar wind interaction is strongest at low solar zenith angles ($SZA$), where the ionosphere experiences a direct head-on impact, and weakens progressively with increasing $SZA$. To account for this dependence, we calculated the flow zenith angle ($SZA_f$), which represents the variation of $P_{SW_p}$ at larger $SZAs$ (see Section~\ref{Solar_wind}). Our analysis shows that during periods of enhanced $P_{SW_p}$, the Venusian ionosphere exhibits either disturbances in the topside structure or the occurrence of a Type 1 bulge.. Figure~\ref{Figure_14} illustrates this variability: under similar solar activity and $SZA$ conditions, the Venusian topside ionosphere responds to high solar wind dynamic pressure ($P_{SW_p}$) by exhibiting either a Type 1 bulge or a disturbance, while the profile with low solar wind dynamic pressure is unaffected. Notably, no clear threshold value of $P_{SW}$ could be identified for the occurrence of the bulge.

%%%%%%%%%%%%%%%%%%%%%%%%%%%%%%%%%%%%%%%%%%%%
%
% Discussion
%
%%%%%%%%%%%%%%%%%%%%%%%%%%%%%%%%%%%%%%%%%%%%
\section{Discussion}
\label{Discussion}
In the present study, we analyze the huge Venus Express VeRa data set, covering both the declining phase of Solar Cycle 23 and the rising phase of Solar Cycle 24. For the first time, we have systematically characterized and studied the bulge of the Venusian ionosphere by analyzing 234 electron density profiles of the VeRa experiment for the dayside ionosphere of Venus. There have been many possible theories of the formation of the bulge in the topside Venusian ionosphere. Most of them are related to the solar wind interaction with the Venusian ionosphere, which has no intrinsic magnetic field. However, the exact mechanism is still unknown.

Previous studies by \citet{bauer1974venus, shinagawa1987one, woo1991magnetization} have proposed that the interaction of the solar wind plays a significant role in generating the bulge observed in the topside Venusian ionosphere. At the time of high solar wind pressure, the interplanetary magnetic field can penetrate into the ionosphere. \citet{bauer1974venus} suggested that this bulge forms at the transition between the photochemical-dominated and solar wind-induced downward diffusion-dominated regions. Supporting this, the 1D MHD model developed by \citet{shinagawa1987one} demonstrated that during periods when the Venusian ionosphere is magnetized due to the high solar wind dynamic pressure \citep{elphic1984venus, luhmann1991magnetic}, a local magnetic field with a maximum and a minimum develops at altitudes near $165~km$ and $205~km$ (because of the solar wind-induced downward vertical convection motion), respectively, and a corresponding electron density bulge appears near $190~km$ altitude.

The peak altitude of the Type 1 bulge morphology (Table~\ref{tab:profiles_table}) is higher than that of the other two morphologies. Our analysis also shows that bulge occurrence is enhanced during periods of low solar activity. Under such conditions, the ionosphere is comparatively weak, with reduced thermal pressure. As a result, it is less able to withstand the solar wind dynamic pressure, allowing the interplanetary magnetic field to penetrate into the ionosphere. With increasing $SZA$, however, the occurrence of the Type 1 morphology declines (Figure~\ref{Figure_6}), consistent with the expectation that solar wind dynamic pressure becomes less effective at larger $SZAs$ due to oblique incidence and reduced momentum transfer to the ionosphere.

A rapid increase in electron temperature at altitudes of $150$–$170~\mathrm{km}$ has been suggested as one possible explanation for the formation of the bulge \citep{fox2006morphology, peter2014dayside}. To date, however, no photochemical model of the Venusian ionosphere has successfully reproduced this feature. \citet{fox2001solar} reported a similar "bulge" near $200~\mathrm{km}$ in their ionospheric model under high solar activity, which they attributed to the high atomic Oxygen (\ce{O}) density in the VTS3 model \citep{hedin1983global} rather than to electron temperature variations. Under high solar activity, the higher solar EUV flux increases the neutral temperature, which in turn increases the neutral scale height and expands the neutral atmosphere \cite{hensley2020dependence}. This expansion alters the composition of the upper atmosphere, raising the \ce{nO/nCO2} ratio at higher altitudes \citep{hedin1983global}. Therefore, a high \ce{nO/nCO2} ratio would predict more frequent bulge appearances during high solar activity. However, our analysis of the VeRa data shows that the bulge is observed more frequently during low solar activity. This suggests that the photochemical processes alone cannot account for the formation of the bulge in the topside Venusian ionosphere.

The precipitation of solar wind into the Venusian ionosphere is another plausible mechanism for bulge formation. Energetic solar wind particles are capable of penetrating to altitudes corresponding to the bulge and depositing a substantial fraction of their energy there \citep{mayyasi2018sporadic}. An initial calculation for solar wind electron precipitation, considering a \ce{CO2}- and \ce{N2}-dominated atmosphere, indicates that electrons with energies between 20~eV and 1~keV can deposit a substantial fraction of their energy within altitudes of $150$–$170~\text{km}$. However, a more comprehensive investigation is required to fully characterize the associated ionization \citep{sheel2012numerical, nakamura2022modeling, gray2025venus}. Detailed modeling by \citet{nakamura2022modeling} suggests that energetic electrons ($<$1~keV) produce maximum ionization above the M2 layer, whereas energetic protons ($>$50~keV) deposit their energy below the M1 layer on the nightside Martian ionosphere. Building on these considerations, the forthcoming Indian \textit{Venus Orbiter Mission} (VOM) is well positioned to investigate solar wind–ionosphere interactions and their role in ionospheric bulge formation through its comprehensive suite of particle and radio instruments. To be launched during the declining phase of Solar Cycle 25, VOM will carry instruments designed to characterize energetic particles and their variability around the Venusian ionosphere. In addition, a topside ionospheric sounder and a radio occultation experiment will provide valuable information on the occurrence, structure, and morphology of bulge features, with the sounder specifically tailored to probe the layered regions above the $\text{V2}$ ionospheric layer. A detailed investigation of the solar wind precipitation is a scope for future work. 

%This analysis shows that during periods of high solar wind dynamic pressure, the electron density profiles often display a bulge-like feature near the V2 peak. With increasing $SZA$, however, both the occurrence and intensity of this feature diminish. This trend is consistent with the expectation that solar wind dynamic pressure is less effective at higher $SZAs$ due to oblique incidence and reduced momentum transfer to the ionosphere.

%%%%%%%%%%%%%%%%%%%%%%%%%%%%%%%%%%%%%%%%%%%%
%
% Conclusions
%
%%%%%%%%%%%%%%%%%%%%%%%%%%%%%%%%%%%%%%%%%%%%
\section{Conclusions}
\label{Conclusions}
Based on the VeRa data from the Venus Express mission, this study provides the first in-depth study of the ionospheric bulge above Venus's main ionospheric peak, focusing on its morphology and key parameters: the peak altitude $h_{m, BL}$ and peak electron density $N_{m, BL}$. The bulge is a persistent feature, observed in over $80\%$ of the analyzed dayside profiles. To enable this comprehensive characterization, we developed an automated, gradient-based routine to systematically identify the bulge in each profile. This method is a significant improvement over previous studies, which relied solely on visual inspection. While the definition of bulge is adapted from \citet{peter2025variability}, our work introduces a new analytical framework for classifying bulge morphology.

The observations show a clear dependence of the bulge's morphology on solar zenith angle ($SZA$). The distinct Type 1 morphology—defined by a secondary electron density maximum above the V2 peak—is most common at low $SZAs$ and during periods of low solar activity. Furthermore, this specific morphology is found to be confined to low latitudes ($\pm40^\circ$). The peak altitude of the bulge, $h_{m, BL}$, shows a negative correlation with $SZA$, likely driven by a decrease in the neutral density at higher altitudes as the terminator approaches. In contrast, the peak electron density of the bulge, $N_{m, BL}$, appears to be largely independent of $SZA$, but is high at high solar activity. Furthermore, the ratio of the bulge peak density to the main peak density $N_{m, BL}/N_{m, V2}$ also increases with rising solar activity, which may be attributed to the expansion of the background neutral atmosphere in response to higher EUV flux.

The dependence of the bulge peak density, altitude, and occurrence rate suggests that its origin cannot be attributed primarily to photochemical processes. Instead, solar wind–related mechanisms, including downward diffusion and particle precipitation, may play a significant role and require further investigation through dedicated model simulations.

% Although a large dataset at low solar activity and low latitudes is put forward. Some insight into the bulge formation, most of the high latitudes in the data are not considered for our analysis. The future Venus Mission from India may add to it, covering a large spatial region.

%%%%%%%%%%%%%%%%%%%%%%%%%%%%%%%%%%%%%%%%%%%%%%%%%%
% CRediT authorship contribution statement
%%%%%%%%%%%%%%%%%%%%%%%%%%%%%%%%%%%%%%%%%%%%%%%%%%
%\section*{CRediT authorship contribution statement}
%\textbf{Satyandra M. Sharma:} Conceptualization, Investigation, Methodology, Writing – original draft. \textbf{Varun Sheel:} Conceptualization, Supervision, Writing – review \& editing. \textbf{Martin Pätzold:} Resources, Writing – review \& editing.

%%%%%%%%%%%%%%%%%%%%%%%%%%%%%%%%%%%%%%%%%%%%%%%%%%
% Declaration of competing interest
%%%%%%%%%%%%%%%%%%%%%%%%%%%%%%%%%%%%%%%%%%%%%%%%%%
\section*{Declaration of competing interest}
The authors declare that they have no known competing financial interests or personal relationships that could have appeared to influence the work reported in this paper.

%%%%%%%%%%%%%%%%%%%%%%%%%%%%%%%%%%%%%%%%%%%%%%%%%%
% Data Availability
%%%%%%%%%%%%%%%%%%%%%%%%%%%%%%%%%%%%%%%%%%%%%%%%%%
\section*{Data Availability}
The VEX-VeRa data radio occultation set is available on request from the experiment team. Contact co-author Pätzold at martin.paetzold@uni-koeln.de.
Publicly available datasets were accessed from the following sources: the solar flux data can be accessed via NASA's OMNIWeb portal at \url{https://omniweb.gsfc.nasa.gov/form/dx1.html}; the ephemerides of Venus are available through the JPL Horizons system at \url{https://ssd.jpl.nasa.gov/horizons/app.html#/}; and the ASPERA-4 solar wind data can be retrieved from the AMDA database at \url{https://amda.irap.omp.eu/index.html}, which requires user login.

%%%%%%%%%%%%%%%%%%%%%%%%%%%%%%%%%%%%%%%%%%%%%%%%%%
% Acknowledgments
%%%%%%%%%%%%%%%%%%%%%%%%%%%%%%%%%%%%%%%%%%%%%%%%%%
\section*{Acknowledgments}
MP thanks the Deutscher Akademischer Austauschdienst (DAAD), grant number 57623043, for supporting his visit to PRL, Ahmedabad, India, and everyone at PRL for their generous and warm hospitality.

%%%%%%%%%%%%%%%%%%%%%%%%%%%%%%%%%%%%%%%%%%%%
%
% Appendix
%
%%%%%%%%%%%%%%%%%%%%%%%%%%%%%%%%%%%%%%%%%%%%
\section*{Appendix}

\subsection*{Morphology of the Bulge}
\label{bulge morphology}

To systematically classify the morphology of the bulge in the topside Venusian ionosphere, we analyzed the gradient of the smoothed electron density profile, denoted as $(N_S)'$. This gradient $(N_S)'$ is computed for each profile using MATLAB “gradient” function, which calculates central differences for interior points and single-sided differences at the edges. The distinct morphological types are defined based on the structure and variation of the $(N_S)'$ in the vicinity of the bulge.

\textbf{TYPE 1 Morphology} is assigned when a positive gradient $(N_S)'$ is detected between the upper boundary of the V2 layer ($h_{BL,1}$) and the altitude of maximum density within the bulge ($h_{m, BL}$). This condition indicates a localized enhancement in electron density, forming a clear bulge above the V2 peak.

\textbf{TYPE 2 and TYPE 3 Morphologies} are determined by comparing the average electron density gradients above and below the bulge peak. Specifically, we compute:
\begin{itemize}
    \item $m_{BL,top}$: the average gradient from $h_{m,BL}$ to the top boundary of the bulge.
    \item $m_{BL,bottom}$: the average gradient from $h_{m,BL}$ to the bottom boundary of the bulge.
\end{itemize}

The morphology is then classified as:
\begin{enumerate}
    \item \textbf{TYPE 2:} if $m_{BL, top} / m_{BL, bottom} > 1$ — indicating a steep increase in density below the bulge, characteristic of a shoulder-like structure.
    \item \textbf{TYPE 3:} if $m_{BL, top} / m_{BL, bottom} < 1$ — indicating a more gradual density decrease, without strong enhancement below the bulge.
\end{enumerate}

This gradient-based classification enables a consistent and reproducible categorization of bulge structures, which may correspond to different physical mechanisms such as localized ionization, thermospheric transport, or interactions with the solar wind.

\printcredits

%% Loading bibliography style file
%\bibliographystyle{model1-num-names}
\bibliographystyle{cas-model2-names}

% Loading bibliography database
\bibliography{cas-refs-Mybib}

% Biography
%\bio{}
% Here goes the biography details.
%\endbio

%\bio{pic1}
% Here goes the biography details.
%\endbio

\end{document}